\documentclass[11pt]{article}
\usepackage{a4wide}
\usepackage{mathrsfs}
\usepackage{latexsym,bm}
\usepackage{graphicx}
\usepackage{indentfirst}
\usepackage{slashed}
\usepackage{amsmath}
\usepackage{amssymb}
\usepackage[usenames,dvipsnames]{color}
\usepackage{hyperref}
\usepackage{epsfig}
\usepackage[titletoc]{appendix}
\usepackage{multirow}%
\usepackage{rotating}
\usepackage[square,numbers,sort&compress]{natbib}
\usepackage{tikz,tikz-feynman,graphbox,subfigure}
\usepackage{ulem}
\usepackage[top=2.9cm,bottom=2.5cm,left=2.8cm,right=3cm]{geometry}
\usepackage{tabularx}
\usepackage{bbm}
\usepackage{lscape}

\newcommand{\email}[1]{\footnote{{\em } \texttt{#1}}}

\newcommand{\bma}{\left(\begin{matrix}}
\newcommand{\ema}{\end{matrix}\right)}

\newcommand{\bra}{\langle}
\newcommand{\ket}{\rangle}
\newcommand{\xpt}{{\chi}{\rm PT}}

\begin{document}

\thispagestyle{empty}
\title{
\Large \bf Axion production in the $\eta\to \pi\pi a$ decay within $SU(3)$ chiral perturbation theory }
\author{\small Jin-Bao Wang$^{a,b}$,\, Zhi-Hui Guo$^{a}$\email{zhguo@hebtu.edu.cn}, \, Zhun Lu$^{b}$\email{zhunlu@seu.edu.cn} ,\,Hai-Qing Zhou$^{b}$\email{zhouhq@seu.edu.cn} \\[0.5em]
{ \small\it ${}^a$ Department of Physics and Hebei Key Laboratory of Photophysics Research and Application, } \\ 
{\small\it Hebei Normal University,  Shijiazhuang 050024, China} 
\\[0.2em] 
{ \small\it ${}^b$ School of Physics, Southeast University, Nanjing 211189, China }
}
\date{}

%

\maketitle
\begin{abstract}
We study the axion and axion-like particle production from the $\eta\to\pi\pi a$ decay within the $SU(3)$ chiral perturbation theory up to the one-loop level. The conventional $SU(3)$ chiral low energy constants are found to be able to reabsorb all the divergences from the chiral loops in the $\eta\to\pi\pi a$ decay amplitude, and hence render the amplitude independent of the renormalization scale. The unitarized $\eta\to\pi\pi a$ decay amplitudes are constructed to take into account the $\pi\pi$ final-state interactions and also properly reproduce the perturbative results from the chiral perturbation theory. Detailed analyses between the perturbative amplitudes and the unitarized ones are given in the phenomenological discussions. By taking the values of the chiral low energy constants in literature, we predict the Dalitz distributions, the spectra of the $\pi\pi$ and $a\pi$ systems, and also the branching ratios of the $\eta\to\pi\pi a$ process by varying $m_a$ from 0 to $m_\eta-2m_{\pi}$. 
\end{abstract}

\section{Introduction}

The hunting of the hypothetical axion ($a$) and axion-like particles (ALPs) constitutes one of the most active research subjects in particle physics nowadays~\cite{Kim:2008hd,Graham:2015ouw,Irastorza:2018dyq,DiLuzio:2020wdo,Choi:2020rgn,Sikivie:2020zpn}. As an elegant resolution of the strong CP problem in QCD, the interaction of the axion and gluons, i.e. $a G^{\mu\nu} \tilde{G}_{\mu\nu}\alpha_s/(8\pi f_a)$, where $G^{\mu\nu}$ is the gluon field strength tensor, $\tilde{G}_{\mu\nu}$ corresponds to its dual and $f_a$ denotes the axion decay constant, is deemed to be the model independent ingredient in the numerous constructions of axion models. In this case, also dubbed as the QCD axion case, not only all the axion interactions but also the axion mass are governed by the single parameter $f_a$~\cite{Weinberg:1977ma}, leading to a stringent constraint on the former parameter. The visible QCD axion scenario, i.e., when $f_a$ lies at the electroweak symmetry breaking scale $v_{EW}\simeq 256$~GeV, proposed in the early pioneer works of Refs.~\cite{Peccei:1977hh,Peccei:1977ur,Weinberg:1977ma,Wilczek:1977pj}, has been ruled out by various experiments~\cite{Kim:1986ax}. As a result, the invisible axion scenario with $f_a\gg v_{EW}$ becomes the focus in various studies. Axion can only feebly interact with particles of the Standard Model (SM), since the interaction strengths of the axion are generally suppressed by $(1/f_a)^n$, being $n$ the number of axions involved. 

In this work, we stick to the model independent term $a G^{\mu\nu} \tilde{G}_{\mu\nu}\alpha_s/(8\pi f_a)$ to account for the axion interaction, and this inevitably introduces the axion couplings with the QCD hadrons in the low energy region. Regarding the axion mass, the $a G\tilde{G}$ term gives a contribution behaving like $m_a^2 \propto m_\pi^2 F_\pi^2/f_a^2$~\cite{Weinberg:1977ma}, which becomes rather tiny in the invisible axion scenario. In the ALP case, it is possible to introduce a bare mass term $m_{a,0}$ to the axion from a UV theory, see the very recent discussions in~Refs.~\cite{DiLuzio:2020wdo,DiLuzio:2016sbl,DiLuzio:2017pfr,Gaillard:2018xgk,Hook:2019qoh,deGiorgi:2024str}, and the axion mass is then disentangled from its decay constant $f_a$. As long as $m_{a,0}\ll f_a$, the presence of the bare mass term $m_{a,0}$ can be considered as a soft breaking of the Peccei-Quinn symmetry. In the phenomenological exploration, $m_{a,0}$ can be taken as a free parameter in the region much below $f_a$. 

The $\eta$ physics constitutes an interesting and important subject at many ongoing and planned large science facilities, such as BESIII~\cite{BESIII:2020nme,Fang:2017qgz}, Jefferson-Lab Eta factory~\cite{Gan:2015nyc,Gan:2017kfr,Somov:2024jiy}, the Rare Eta Decays To Probe New Physics (REDTOP) experiment~\cite{Gatto:2019dhj,REDTOP:2022slw}, the Super Tau-Charm Facility (STCF)~\cite{Achasov:2023gey}, etc. The large data samples of $\eta$ at such facilities can provide precious experimental opportunities to explore the axion production mechanism in the $\eta\to\pi\pi a$ decay. In this study, we will concentrate on the axion and ALP production from the $\eta\to\pi\pi a$ process by varying the ALP mass from zero to $m_\eta-2m_{\pi}$.
The leading order (LO) amplitudes of $\eta^{(')}\to\pi\pi a$ in a $U(3)$ chiral framework have been calculated in Ref.~\cite{Gan:2020aco}, and it turns out that apart from the tiny isospin breaking corrections there are only constant terms, which are proportional to $m_\pi^2$, remaining in the amplitudes. A variant chiral calculation in Ref.~\cite{Landini:2019eck} gives comparable expressions as the LO results in Ref.~\cite{Gan:2020aco} that only contain the constant terms,  and the energy dependent parts are absent at this order. Therefore this implies that the higher order effects could give important contributions, comparing with the LO ones. The $\eta^{(')}\to\pi\pi a$ decay widths are calculated in Ref.~\cite{Alves:2020xhf} by including the bare resonance exchanges at tree level. Very recently, the $\eta^{(')}\to\pi\pi a$ amplitudes are calculated by incorporating the $\pi\pi$ final state interaction into the LO $\eta^{(')}\to\pi\pi a$ amplitudes in Ref.~\cite{Alves:2024dpa}. 
In the current study, we pursue the $\eta\to\pi\pi a$ calculation by working out the complete one-loop expressions in the $SU(3)$ chiral perturbation theory ($\chi$PT) and study the unitarization effects, which are then employed to explore the phenomenological consequences, such as the Dalitz decay distributions, the spectra of the $\pi\pi$ and $a\pi$ systems, and the $\eta\to\pi\pi a$ decay widths as a function of the ALP mass.

This paper is organized as follows. In Sec.~\ref{sec.alplag}, the axion $SU(3)$ chiral Lagrangian up to $\mathcal{O}(p^4)$ is elaborated. In Sec.~\ref{sec.mixing}, we discuss the ALP-$\eta$ mixing and the ALP mass up to one-loop level. The decay amplitude of the $\eta\to\pi\pi a$ process is calculated up to $\mathcal{O}(p^4)$ in Sec.~\ref{sec.decayamp}, where the unitarization procedure is also discussed to include the $\pi\pi$ final-state interaction. In Sec.~\ref{sec.pheno}, we study the phenomenologies in the $\eta\to\pi\pi a$ decay by separately analyzing the perturbative $\xpt$ amplitudes and the unitarized ones. We give a short summary and conclusions in Sec.~\ref{sec.concl}.

\section{Axion $SU(3)$ chiral Lagrangian}\label{sec.alplag}

Although it is straightforward to include additional model-dependent unknown axion couplings with quarks, we carry out the calculations by taking the model-independent axion interaction, namely the couplings exclusively derived from the $a G^{\mu\nu} \tilde{G}_{\mu\nu}\alpha_s/(8\pi f_a)$ operator. In this way it allows us to proceed the phenomenological discussions later with more definite results, since all the unknown parameters except $f_a$ have been determined in the $\xpt$ studies with light-flavor mesons~\cite{Bijnens:2014lea}. 

A practical way to handle the $a G^{\mu\nu} \tilde{G}_{\mu\nu}\alpha_s/(8\pi f_a)$ operator is to perform an axial transformation of the quark fields
\begin{equation}\label{eq.axialtrans}
q \to \exp\left(i\frac{a}{2f_a}Q_a\gamma_5\right)q\,,
\end{equation}
where $Q_a$ is a $3\times3$ hermitian matrix in light-quark flavor space with its trace $\langle Q_a \rangle=1$. In such a manner one can eliminate the $aG\widetilde{G}$ term and meanwhile introduces two new modifications, namely the axion-quark derivative interaction and the modified quark mass term. The resulting effective axion Lagrangian after the transformation of Eq.~\eqref{eq.axialtrans} reads 
\begin{equation}\label{eq.LALP}
\mathcal{L}_{\text{ALP}}=\frac{1}{2}\partial_{\mu}a\partial^{\mu}a-\frac{1}{2}m_{a,0}^2a^2
-\frac{\partial_{\mu}a}{2f_a}\bar{q}\gamma^{\mu}\gamma_5Q_aq
-\bar{q}\exp\left(i\frac{a}{2f_a}Q_a\gamma_5\right)M_q\exp\left(i\frac{a}{2f_a}Q_a\gamma_5\right)q\,,
\end{equation}
where $m_{a,0}$ denotes the bare ALP mass and the light-flavor quark mass matrix is given by $M_q={\rm diag}(m_u,m_d,m_s)$. It is verified that the strong isospin-breaking correction from the $\epsilon_I\equiv B_0(m_d-m_u)$ factor, with $B_0$ proportional to the quark condensate, in the $\eta \to \pi \pi a$ decay begins at $\mathcal{O}(\epsilon_I^2)$, which therefore can be safely neglected. As a result we will work in isospin limit and take $M_q=\text{diag}(\hat{m},\,\hat{m},\,m_s)$ in later discussions, being $\hat{m}=(m_u+m_d)/2$.   

Next we match the effective axion Lagrangian in Eq.~(\ref{eq.LALP}) to the  $SU(3)~\xpt$. The LO chiral Lagrangian with axion can be constructed as
\begin{equation}\label{eq.L2}
\mathcal{L}_2=
\frac{F^2}{4}\langle\partial_{\mu}U\partial^{\mu}U^{\dagger}+\chi_aU^{\dagger}+U\chi_a^{\dagger}\rangle
+\frac{\partial_{\mu}a}{2f_a}J_A^{\mu}\big|_{\mathrm{LO}}+ \frac{1}{2}\partial_{\mu}a\partial^{\mu}a-\frac{1}{2}m_{a,0}^2a^2\,,
\end{equation}
where the octet of pseudo Nambu-Goldstone bosons (pNGBs) is collected in $U={\exp}\big(i\Phi/F\big)$ with
\begin{equation}\label{phi1}
\Phi \,=\, 
\left(\begin{matrix}
\pi^0+\frac{1}{\sqrt{3}}\eta_8&\sqrt{2}\pi^+&\sqrt{2}K^+\\
\sqrt{2}\pi^-&-\pi^0+\frac{1}{\sqrt{3}}\eta_8&\sqrt{2}K^0\\
\sqrt{2}K^-&\sqrt{2}\bar{K}^0&-\frac{2}{\sqrt{3}}\eta_8
\end{matrix}\right)\,.
\end{equation}
It is noted that the physical $\eta$ meson is approximated by the $SU(3)$ octet $\eta_8$ field in the $SU(3)$ $\xpt$.
The axion dressed quark mass matrix is given by  $\chi_a=2B_0M(a)$ with
\begin{equation}
M(a)\equiv\exp\left(-i\frac{a}{2f_a}Q_a\right)M_q\exp\left(-i\frac{a}{2f_a}Q_a\right)\,.
\end{equation} 
There is arbitrariness for the choice of $Q_a$ in Eq.~\eqref{eq.axialtrans} with the constraint $\bra Q_a\ket =1$, though different choices do not change the physical results. It is customary to take $Q_a=M_q^{-1}/\langle{M_q^{-1}}\rangle$~\cite{Georgi:1986df} which eliminates the mass mixing between ALP and octet pNGBs at LO. 
The $J_A^{\mu}\big|_{\mathrm{LO}}$ term in Eq.~(\ref{eq.L2}) corresponds to the hadronic parameterization of the quark axial-vector current $\bar{q}\gamma^{\mu}\gamma_5(-Q_a)q$ in Eq.~(\ref{eq.LALP}) at LO. The explicit expression of $J_A^{\mu}\big|_{\mathrm{LO}}$ is given by 
\begin{equation}\label{eq.jalo}
J_A^{\mu}\big|_{\text{LO}}=-i\frac{F^2}{2}\langle Q_a\left\{\partial^{\mu}U,\,U^{\dagger}\right\} \rangle\,.
\end{equation}
$Q_a$ contains both the flavor-octet and flavor-singlet components. It is easy to verify that the singlet component of the LO hadronic axial-vector current in Eq.~\eqref{eq.jalo} vanishes and only the octet current contributes to the derivative axion couplings at this order.

At next-to-leading order (NLO) in the chiral expansion, a physical amplitude can receive contributions both from the one-loop Feynman diagrams and the local operators of the $\mathcal{O}(p^4)$ chiral Lagrangian~\cite{Gasser:1984gg}
\begin{equation}\label{eq.L4}
\begin{aligned}
\mathcal{L}_{4}\supset&
L_1\langle\partial_{\mu}U\partial^{\mu}U^{\dagger}\rangle\langle\partial_{\nu}U\partial^{\nu}U^{\dagger}\rangle
+L_2\langle\partial_{\mu}U\partial_{\nu}U^{\dagger}\rangle\langle\partial^{\mu}U\partial^{\nu}U^{\dagger}\rangle
+L_3\langle\partial_{\mu}U\partial^{\mu}U^{\dagger}\partial_{\nu}U\partial^{\nu}U^{\dagger}\rangle
\\
&+L_4\langle\partial_{\mu}U\partial^{\mu}U^{\dagger}\rangle\langle\chi_aU^{\dagger}+U\chi_a^{\dagger}\rangle
+L_5\langle\partial_{\mu}U\partial^{\mu}U^{\dagger}(\chi_aU^{\dagger}+U\chi_a^{\dagger})\rangle
+L_6\left(\langle\chi_aU^{\dagger}+U\chi_a^{\dagger}\rangle\right)^2
\\ &
+L_7\left(\langle\chi_aU^{\dagger}-U\chi_a^{\dagger}\rangle\right)^2
+L_8\langle\chi_aU^{\dagger}\chi_aU^{\dagger}+U\chi_a^{\dagger}U\chi_a^{\dagger}\rangle
+\frac{\partial_{\mu}a}{2f_a}J^{\mu}_A\big|_{\text{NLO}}\,, 
\end{aligned}
\end{equation}
with the NLO hadronic axial-vector current 
\begin{equation}
\begin{aligned}\label{eq.janlo}
J^{\mu}_A\big|_{\text{NLO}}=&
-4iL_1\langle\bar{Q}_a\{U^{\dagger},\,\partial^{\mu}U\}\rangle\langle\partial_{\nu}U\partial^{\nu}U^{\dagger}\rangle
\\
&-2iL_2\langle\bar{Q}_a\{U^{\dagger},\,\partial_{\nu}U\}\rangle
\langle\partial^{\mu}U\partial^{\nu}U^{\dagger}+\partial^{\nu}U\partial^{\mu}U^{\dagger}\rangle
\\
&-2iL_3\langle\partial^{\mu}U\{\bar{Q}_a,\,U^{\dagger}\}\partial_{\nu}U\partial^{\nu}U^{\dagger}\rangle
+2iL_3\langle\{\bar{Q}_a,\,U\}\partial^{\mu}U^{\dagger}\partial_{\nu}U\partial^{\nu}U^{\dagger}\rangle
\\
&-2iL_4\langle\bar{Q}_a\{U^{\dagger},\,\partial^{\mu}U\}\rangle\langle\chi U^{\dagger}+U\chi^{\dagger}\rangle
\\
&-iL_5\langle\partial^{\mu}U\{\bar{Q}_a,\,U^{\dagger}\}(\chi U^{\dagger}+U\chi^{\dagger})\rangle
+iL_5\langle\{\bar{Q}_a,\,U\}\partial^{\mu}U^{\dagger}(\chi U^{\dagger}+U\chi^{\dagger})\rangle\,,
\end{aligned}
\end{equation}
where $\bar{Q}_a$ denotes the octet component of $Q_a$. We have only shown the relevant operators to this work in Eqs.~(\ref{eq.L4}) and (\ref{eq.janlo}). Although the singlet component of the hadronic axial current at LO in Eq.~\eqref{eq.jalo} vanishes, it does contribute at NLO in Eq.~\eqref{eq.janlo}. Nevertheless, a more complete framework to investigate the singlet axial-vector current would be the $U(3)$ $\xpt$ that explicitly includes the QCD $U_A(1)$ anomaly effect and the iso-singlet $\eta_0$ state, and this is beyond the scope of the present study that relies on the $SU(3)$ $\xpt$. In fact the inclusion of the singlet axial current in the $U(3)$ $\xpt$ with the explicit $\eta_0$ field will change the renormalization of the $SU(3)$ $\xpt$ LECs~\cite{Herrera-Siklody:1996tqr,Kaiser:2000gs}. Since we stick to the $SU(3)$ framework to calculate the $\eta\to \pi\pi a$ process, the singlet component for the NLO hadronic axial currents in Eq.~\eqref{eq.janlo} will be omitted as an approximation. In this case, the $\eta\to\pi\pi a$ amplitude can be renormalized with the conventional renormalization conditions for the standard $SU(3)$ $\xpt$ LECs~\cite{Gasser:1984gg}, as will be demonstrated later.

\section{ALP-$\eta$ mixing and ALP mass up to NLO}\label{sec.mixing}

For the perturbative calculation of the amplitudes, we need to first solve the mixing of the ALP and $\pi^0$, $\eta_8$ fields. The key difference between the ALP-$\pi^0$ and ALP-$\eta_8$ mixing is that the former vanishes in the isospin limit and the latter remains finite. It is pointed out that the ALP-$\pi^0$ mixing will contribute to the $\eta\to \pi\pi a$ amplitude at the level of $\mathcal{O}(\epsilon_I^2)$, and we will neglect the strong isospin breaking effects throughout. In this section, we deal with ALP-$\eta_8$ mixing up to NLO in the $SU(3)$ $\xpt$ and calculate the NLO chiral correction to the ALP mass as well. 
The chiral correction to the ALP mass comes from the couplings between ALP and pNGBs and thus is of $\mathcal{O}(1/f_a^2)$.
Keeping the terms up to $\mathcal{O}(1/f_a^2)$, the LO quadratic terms of ALP and $\eta_8$ fields from LO Lagrangian Eq.~(\ref{eq.L2}) take the form
\begin{equation}
\mathcal{L}_{\mathrm{mix}}^{\mathrm{LO}}=\frac{1}{2}\partial_{\mu}a\partial^{\mu}a+\frac{1}{2}\partial_{\mu}\eta_8\partial^{\mu}\eta_8+\frac{F}{f_a} C_k^{a\eta}\partial_{\mu}a\partial^{\mu}\eta_8-\frac{1}{2}\left(m_{a,0}^2+\frac{F^2}{f_a^2}C_m^a\right)a^2-\frac{1}{2}\bar{m}_{\eta_8}^2\eta_8^2\,,\label{eq.LmixLO}
\end{equation}
with
\begin{equation}
C_k^{a\eta}=\frac{\bar{m}_{\eta_8}^2-\bar{m}_{\pi}^2}{2\sqrt{3}\bar{m}_{\eta_8}^2}\,,
\quad
C_m^a=\frac{\bar{m}_{\pi}^2(3\bar{m}_{\eta_8}^2-\bar{m}_{\pi}^2)}{12\bar{m}_{\eta_8}^2}\,.
\end{equation}
$\bar{m}_{\pi}$ and $\bar{m}_{\eta_8}$ are the LO masses of $\pi$ and $\eta_8$ contributed by light-quark masses derived from Eq.~(\ref{eq.L2})
\begin{equation}
\bar{m}_{\pi}^2=2B_0\hat{m}\,,
\quad
\bar{m}_{\eta_8}^2=\frac{2B_0}{3}(\hat{m}+2m_s)\,,
\end{equation}
and the LO kaon mass is $\bar{m}_K^2=B_0(\hat{m}+m_s)$, which gives the celebrated Gell-Mann-Okubo relation $4\bar{m}_K^2=3\bar{m}_{\eta_8}^2+\bar{m}_{\pi}^2.$ Since we are interested in $\eta\to\pi\pi a$ decay, for convenience, we will express the light-quark masses $\hat{m}$ and $m_s$ in terms of $\bar{m}_{\pi}^2$ and $\bar{m}_{\eta_8}^2$ in this work.

By performing the following field redefinitions for ALP and $\eta_8$ fields
\begin{align}
&a=\left[1+\frac{1}{2}{(C_k^{a\eta})}^2\frac{m_{a,0}^2(m_{a,0}^2-2\bar{m}_{\eta_8}^2)}{(m_{a,0}^2-\bar{m}_{\eta_8}^2)^2}\frac{F^2}{f_a^2}\right]\widetilde{a}-\frac{F}{f_a} C_k^{a\eta}\frac{\bar{m}_{\eta_8}^2}{\bar{m}_{\eta_8}^2-m_{a,0}^2}\widetilde{\eta}\,,\label{eq.LOredef.a}
\\
&\eta_8=\left[1+\frac{1}{2}{(C_k^{a\eta})}^2\frac{\bar{m}_{\eta_8}^2(\bar{m}_{\eta_8}^2-2m_{a,0}^2)}{(m_{a,0}^2-\bar{m}_{\eta_8}^2)^2}\frac{F^2}{f_a^2}\right]\widetilde{\eta}-\frac{F}{f_a} C_k^{a\eta}\frac{m_{a,0}^2}{m_{a,0}^2-\bar{m}_{\eta_8}^2}\widetilde{a}\,,\label{eq.LOredef.eta}
\end{align}
the LO quadratic terms Eq.~(\ref{eq.LmixLO}) can be diagonalized to the canonical form up to $\mathcal{O}(1/f_a^2)$
\begin{equation}
\mathcal{L}_{\mathrm{mix}}^{\mathrm{LO}}=\frac{1}{2}\partial_{\mu}\widetilde{a}\partial^{\mu}\widetilde{a}-\frac{1}{2}\bar{m}_a^2\widetilde{a}^2+\frac{1}{2}\partial_{\mu}\widetilde{\eta}\partial^{\mu}\widetilde{\eta}-\frac{1}{2}\bar{m}_{\eta}^2\widetilde{\eta}^2\,.\label{eq.LmixLOdiag}
\end{equation}
The resulting LO masses of ALP and $\eta$ after the elimination of the mixing read 
\begin{eqnarray}\label{eq.mabar}
\bar{m}_a^2&=&m_{a,0}^2+\frac{F^2}{f_a^2}\left(C_m^a+\frac{{(C_k^{a\eta})}^2 m_{a,0}^4}{m_{a,0}^2-\bar{m}_{\eta_8}^2}\right)\,, \\
\bar{m}_{\eta}^2&=&\bar{m}_{\eta_8}^2+\frac{F^2}{f_a^2}\frac{{(C_k^{a\eta})}^2\bar{m}_{\eta_8}^4}{\bar{m}_{\eta_8}^2-m_{a,0}^2}\,.
\end{eqnarray}  
By keeping the terms up to $\mathcal{O}(1/f_a^2)$, we can replace $m_{a,0}$ in Eqs.~(\ref{eq.LOredef.a}) and (\ref{eq.LOredef.eta}) by $\bar{m}_a$
\begin{align}
&a=\left[1+\frac{1}{2}{(C_k^{a\eta})}^2\frac{\bar{m}_{a}^2(\bar{m}_{a}^2-2\bar{m}_{\eta_8}^2)}{(\bar{m}_{a}^2-\bar{m}_{\eta_8}^2)^2}\frac{F^2}{f_a^2}\right]\widetilde{a}-\frac{F}{f_a} C_k^{a\eta}\frac{\bar{m}_{\eta_8}^2}{\bar{m}_{\eta_8}^2-\bar{m}_{a}^2}\widetilde{\eta}+\mathcal{O}(1/f_a^3)\,,\label{eq.LOredef2.a}
\\
&\eta_8=\left[1+\frac{1}{2}{(C_k^{a\eta})}^2\frac{\bar{m}_{\eta_8}^2(\bar{m}_{\eta_8}^2-2\bar{m}_{a}^2)}{(\bar{m}_{a}^2-\bar{m}_{\eta_8}^2)^2}\frac{F^2}{f_a^2}\right]\widetilde{\eta}-\frac{F}{f_a} C_k^{a\eta}\frac{\bar{m}_{a}^2}{\bar{m}_{a}^2-\bar{m}_{\eta_8}^2}\widetilde{a}+\mathcal{O}(1/f_a^3)\,,\label{eq.LOredef2.eta}
\end{align}
due to the fact $\bar{m}_a^2=m_{a,0}^2+\mathcal{O}(1/f_a^2)$.  

The $\widetilde{a}$ and $\widetilde{\eta}$ fields in Eq.~(\ref{eq.LmixLOdiag}) are now in the canonical form at LO. However, the $\widetilde{a}$ and $\widetilde{\eta}$ fields will get mixed again at NLO. Define the two-point one-particle-irreducible (1PI) amplitudes for ALP and $\eta$ fields as $\Sigma_{aa}(p^2)$, $\Sigma_{a\eta}(p^2)$, $\Sigma_{\eta\eta}(p^2)$, whose explicit expressions at the one-loop level are provided in Appendix~\ref{appendix.A}.  Up to $\mathcal{O}(p^4)$ they can be decomposed as
$\Sigma_{ij}^{(4)}(p^2)=p^2\Sigma_{ij}^{(4)'}(0)+\Sigma_{ij}^{(4)}(0)$ with $\Sigma_{ij}^{(4)'}(0)=\mathrm{d}\Sigma_{ij}^{(4)}(p^2)/\mathrm{d}p^2 |_{p^2=0}$, with $i,j=a,\eta$. In fact, since $\mathrm{d}\Sigma_{ij}^{(4)}(p^2)/\mathrm{d}p^2$ does not depend on $p^2$ at the $\mathcal{O}(p^4)$ level, we will simply denote $\Sigma_{ij}^{(4)'}(0)$ as $\Sigma_{ij}^{(4)'}$ in the following discussions. The number $n$ inside the bracket in the superscript denotes the chiral order of the amplitude. The quadric form of the ALP and $\eta$ fields up to NLO can be written as
\begin{equation}
\begin{aligned}
\mathcal{L}_{\mathrm{mix}}^{\mathrm{NLO}}=&
\frac12(1-\Sigma_{aa}^{(4)'})\partial_{\mu}\widetilde{a}\partial^{\mu}\widetilde{a}
+\frac12(1-\Sigma_{\eta\eta}^{(4)'})\partial_{\mu}\widetilde{\eta}\partial^{\mu}\widetilde{\eta}
-\Sigma_{a\eta}^{(4)'}\partial_{\mu}\widetilde{a}\partial^{\mu}\widetilde{\eta}
\\
&-\frac12\left[\bar{m}_a^2+\Sigma_{aa}^{(4)}(0)\right]\widetilde{a}^2
-\frac12\left[\bar{m}_{\eta}^2+\Sigma_{\eta\eta}^{(4)}(0)\right]\widetilde{\eta}^2
-\Sigma_{a\eta}^{(4)}(0)\widetilde{a}\widetilde{\eta}\,.
\end{aligned}
\end{equation}
We first eliminate the kinetic mixing up to NLO with the field redefinitions
\begin{equation}
\widetilde{a}=\left(1+\frac12\Sigma_{aa}^{(4)'}\right)\hat{a}+\frac12\Sigma_{a\eta}^{(4)'}\hat{\eta}\,,
\quad
\widetilde{\eta}=\left(1+\frac12\Sigma_{\eta\eta}^{(4)'}\right)\hat{\eta}+\frac12\Sigma_{a\eta}^{(4)'}\hat{a}\,.\label{eq.fieldshat}
\end{equation}
By using the field variables $\hat{a}$ and $\hat{\eta}$, the kinetic terms can be diagonalized to the canonical form 
\begin{equation} \label{eq.nlomixhat}
\mathcal{L}_{\mathrm{mix}}^{\mathrm{NLO}}=
\frac12\partial_{\mu}\hat{a}\partial^{\mu}\hat{a}
+\frac12\partial_{\mu}\hat{\eta}\partial^{\mu}\hat{\eta}
-\frac12 \hat{m}_a^2\hat{a}^2-\frac12\hat{m}_{\eta}^2\hat{\eta}^2
-\left[\frac{1}{2}(\bar{m}_a^2+\bar{m}_{\eta}^2)\Sigma_{a\eta}^{(4)'}+\Sigma_{a\eta}^{(4)}(0)\right]\hat{a}\hat{\eta}\,,
\end{equation}
with 
\begin{align}
&\hat{m}_a^2=\bar{m}_a^2+\Sigma_{aa}^{(4)'}\bar{m}_a^2+\Sigma_{aa}^{(4)}(0)=\bar{m}_a^2+\Sigma_{aa}^{(4)}(\bar{m}_a^2)\,,\label{eq.ma}
\\
&\hat{m}_{\eta}^2=\bar{m}_{\eta}^2+\Sigma_{\eta\eta}^{(4)'}\bar{m}_{\eta}^2+\Sigma_{\eta\eta}^{(4)}(0)=\bar{m}_{\eta}^2+\Sigma_{\eta\eta}^{(4)}(\bar{m}_{\eta}^2)\,.\label{eq.meta}
\end{align}
Next we deal with the $\hat{a}$-$\hat{\eta}$ mass mixing via a perturbative real orthogonal transformation
\begin{equation}\label{eq.aetahatrel}
\hat{a}=a_{\mathrm{phy}}+\delta_m \, \eta_{\mathrm{phy}}\,,
\quad
\hat{\eta}=\eta_{\mathrm{phy}}-\delta_m\, a_{\mathrm{phy}}\,,
\end{equation}
with
\begin{equation}
\delta_m =-\frac{\Sigma_{a\eta}^{(4)'}\frac{1}{2}(\bar{m}_a^2+\bar{m}_{\eta}^2)+\Sigma_{a\eta}^{(4)}(0)}{\hat{m}_a^2-\hat{m}_{\eta}^2}\,.\label{eq.fieldsphy}
\end{equation}
Combining Eqs.~(\ref{eq.fieldshat}) and (\ref{eq.aetahatrel}) one can get the relations between field variables $\widetilde{a}$, $\widetilde{\eta}$ and the NLO renormalized fields $a_{\mathrm{phy}}$ and $\eta_{\mathrm{phy}}$, 
\begin{equation}
\widetilde{a}=\left(1+\frac{1}{2}\Sigma_{aa}^{(4)'}\right)a_{\mathrm{phy}}+\frac{\Sigma_{a\eta}^{(4)}(m_{\eta}^2)}{m_{\eta}^2-m_a^2}\eta_{\mathrm{phy}}\,,
\quad
\widetilde{\eta}=\left(1+\frac12\Sigma_{\eta\eta}^{(4)'}\right)\eta_{\mathrm{phy}}+\frac{\Sigma_{a\eta}^{(4)'}(m_a^2)}{m_a^2-m_{\eta}^2}a_{\mathrm{phy}}\,.\label{eq.FieldsRenormalization}
\end{equation}
Up to NLO $a_{\mathrm{phy}}$ and $\eta_{\mathrm{phy}}$ become the canonical fields 
\begin{equation}\label{eq.nlomixphy}
\mathcal{L}_{\mathrm{mix}}^{\mathrm{NLO}}=
\frac12\partial_{\mu}a_{\mathrm{phy}}\partial^{\mu}a_{\mathrm{phy}}-\frac12m_a^2a_{\mathrm{phy}}^2+\frac12\partial_{\mu}\eta_{\mathrm{phy}}\partial^{\mu}\eta_{\mathrm{phy}}-\frac12m_{\eta}^2\eta_{\mathrm{phy}}^2\,,
\end{equation}
where $m_a$ and $m_{\eta}$ denote the physical masses of the ALP and $\eta$. It is straightforward to verify that the mixing terms in the field redefinitions~\eqref{eq.aetahatrel} only affect the masses of ALP and $\eta$ at the level of $\mathcal{O}(p^6)$. As a result, the physical masses $m_a$ and $m_{\eta}$ in Eq.~\eqref{eq.nlomixphy} simply equal to $\hat{m}_a$ and $\hat{m}_{\eta}$ in Eq.~\eqref{eq.nlomixhat}, respectively. The explicit expression of the physical ALP mass up to NLO takes the form
\begin{eqnarray}\label{eq.maphy}
m_a^2=\bar{m}_a^2+\Sigma_{aa}^{(4)}(\bar{m}_a^2)\,,
\end{eqnarray}
where the LO ALP mass $\bar{m}_a^2$ is given in Eq.~\eqref{eq.mabar} and the NLO chiral correction reads 
\begin{equation}
\begin{aligned}
\Sigma_{aa}^{(4)}(\bar{m}_a^2)=
&\frac{1}{f_a^2}\Bigg\{
\frac{2L_6^rm_{\pi}^2(3m_{\eta}^2-m_{\pi}^2)(m_{\eta}^2+m_{\pi}^2)}{m_{\eta}^2}
+\frac{2(3L_7^r+L_8^r)m_{\pi}^4(3m_{\eta}^2-m_{\pi}^2)^2}{3m_{\eta}^4}
\\
&-\frac{m_{\pi}^2(3m_{\eta}^2-m_{\pi}^2)^2}{24m_{\eta}^4}\frac{m_{\pi}^2}{(4\pi)^2}\log\frac{m_{\pi}^2}{\mu^2}-\frac{m_{\pi}^2(9m_{\eta}^4-m_{\pi}^4)}{36m_{\eta}^4}\frac{m_{K}^2}{(4\pi)^2}\log\frac{m_{K}^2}{\mu^2}
\\
&-\frac{m_{\pi}^2(3m_{\eta}^2-m_{\pi}^2)(m_{\eta}^2+m_{\pi}^2)}{72m_{\eta}^4}\frac{m_{\eta}^2}{(4\pi)^2}\log\frac{m_{\eta}^2}{\mu^2}
\Bigg\}
\\
&+\frac{X}{f_a^2}\Bigg\{
\frac{4\sqrt{3}L_4^r\bar{m}_a^2(m_{\eta}^4-m_{\pi}^4)}{m_{\eta}^2}
+\frac{8L_5^r\bar{m}_a^2(m_{\eta}^2-m_{\pi}^2)}{\sqrt{3}}
\\
&+\frac{8\sqrt{3}(3L_7^r+L_8^r)m_{\pi}^2(3m_{\eta}^4-4m_{\eta}^2m_{\pi}^2+m_{\pi}^4)}{3m_{\eta}^2}
\\
&+\frac{m_{\pi}^2(3m_{\eta}^2-m_{\pi}^2)}{2\sqrt{3}m_{\eta}^2}\frac{m_{\pi}^2}{(4\pi)^2}\log\frac{m_{\pi}^2}{\mu^2}
-\frac{3m_{\eta}^2m_{\pi}^2-m_{\pi}^4}{6\sqrt{3}m_{\eta}^2}\frac{m_{\eta}^2}{(4\pi)^2}\log\frac{m_{\eta}^2}{\mu^2}
\\
&-\frac{\left[3m_{\eta}^2m_{\pi}^2-m_{\pi}^4+6\bar{m}_a^2(m_{\eta}^2-m_{\pi}^2)\right]}{3\sqrt{3}m_{\eta}^2}\frac{m_{K}^2}{(4\pi)^2}\log\frac{m_{K}^2}{\mu^2}
\Bigg\}
\\
&+\frac{X^2}{f_a^2}\Bigg\{
-12L_4^r\bar{m}_a^2(m_{\eta}^2+m_{\pi}^2)-8L_5^r\bar{m}_a^2m_{\eta}^2+24L_6^rm_{\eta}^2(m_{\eta}^2+m_{\pi}^2)
\\
&+24L_7^r(m_{\eta}^2-m_{\pi}^2)^2+8L_8^r(3m_{\eta}^4-2m_{\eta}^2m_{\pi}^2+m_{\pi}^4)
-\frac{m_{\pi}^2}{2}\frac{m_{\pi}^2}{(4\pi)^2}\log\frac{m_{\pi}^2}{\mu^2}
\\
&+\frac{3\bar{m}_a^2+m_{\pi}^2}{3}\frac{m_{K}^2}{(4\pi)^2}\log\frac{m_K^2}{\mu^2}
-\frac{(4m_{\eta}^2-m_{\pi}^2)}{6}\frac{m_{\eta}^2}{(4\pi)^2}\log\frac{m_{\eta}^2}{\mu^2}
\Bigg\}\,,\label{eq.Sigma_aa}
\end{aligned}
\end{equation}
with $X=C_k^{a\eta}\frac{\bar{m}_a^2}{\bar{m}_a^2-\bar{m}_{\eta_8}^2}$. In the QCD axion scenario with $m_{a,0}^2=0$, the LO mass of QCD axion is given by $\bar{m}_a^2=(F^2/f_a^2)C_m^a$. Then, the last two terms proportional to $X$ and $X^2$ in the expression of $\Sigma_{aa}^{(4)}(\bar{m}_a^2)$ in Eq.~\eqref{eq.Sigma_aa} can be ignored up to $\mathcal{O}(1/f_a^2)$, and thus one can just keep the first term in Eq.~(\ref{eq.Sigma_aa}) when considering the QCD axion case. The NLO correction to the QCD axion mass given here is in consistent with the result given in Ref.~\cite{Lu:2020rhp} in the isospin limit. 
In the following, we will use the NLO renormalized canonical fields $a_{\mathrm{phy}}$ and $\eta_{\mathrm{phy}}$ to perform the perturbative calculation for the $\eta\to\pi\pi a$ decay.

\section{$\eta\to\pi\pi a$ decay amplitude up to NLO and its unitarization}\label{sec.decayamp}

We are interested in the $\eta\to\pi\pi a$ decay with the ALP mass $m_a$ varying from $0$ to $m_{\eta}-2m_{\pi}$. 
The Mandelstam variables to describe the kinematics of the $\eta(P_{\eta}) \to \pi(p_1)\pi(p_2)a(p_a)$ decay are defined as 
\begin{equation}
\begin{aligned}
&s=(P_{\eta}-p_a)^2=(p_1+p_2)^2=m_{\pi\pi}^2\,,
\\
&t=(P_{\eta}-p_1)^2=(p_a+p_2)^2=m_{a\pi}^2\,,
\\
&u=(P_{\eta}-p_2)^2=(p_a+p_1)^2\,,
\end{aligned}
\end{equation}
which satisfy the constraint $s+t+u=m_{\eta}^2+m_a^2+2m_{\pi}^2$.
We consider the $\eta\to\pi\pi a$ decay at the level of $\mathcal{O}(1/f_a)$ and calculate the chiral corrections of the decay amplitudes up to $\mathcal{O}(p^4)$. According to Eq.~\eqref{eq.maphy}, the difference between the physical ALP mass $m_a$ and the bare one $m_{a,0}$ lies at the level of $\mathcal{O}(1/f_a^2)$, i.e. $m_a^2=m_{a,0}^2+\mathcal{O}(1/f_a^2)$, which indicates that the replacement of $m_{a,0}$ with $m_a$ in the $\eta\to\pi\pi a$ amplitude will only give effects at $\mathcal{O}(1/f_a^3)$, beyond the $\mathcal{O}(1/f_a)$ precision we are considering in this work. Therefore, in the following we do not need to distinguish the bare mass $m_{a,0}$ and the physical mass $m_a$ for the ALP, both of which will be simply denoted as $m_a$. 

\begin{figure}[t]
	\subfigure[\label{fig.1a}]{\includegraphics[width=0.24\textwidth]{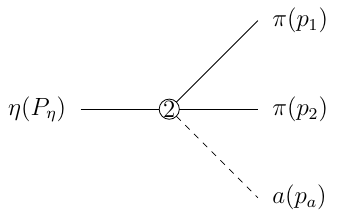}}
	\subfigure[\label{fig.1b}]{\includegraphics[width=0.24\textwidth]{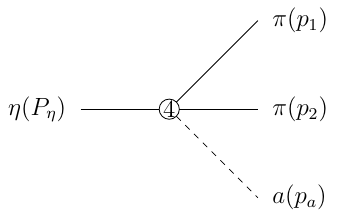}}
	\subfigure[\label{fig.1c}]{\includegraphics[width=0.24\textwidth]{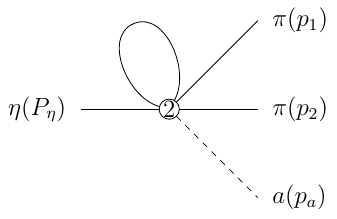}}
	\subfigure[\label{fig.1d}]{\includegraphics[width=0.24\textwidth]{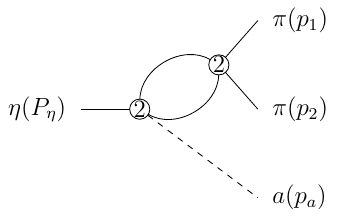}}
	\subfigure[\label{fig.1e}]{\includegraphics[width=0.24\textwidth]{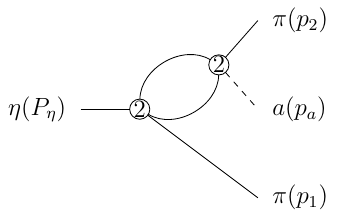}}
	\subfigure[\label{fig.1f}]{\includegraphics[width=0.24\textwidth]{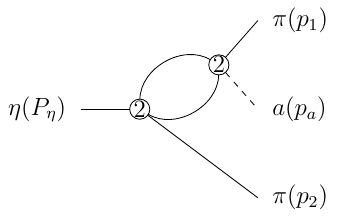}}
	\caption{Feynman diagrams for the $\eta\to\pi\pi a$ decay up to NLO. The numbers in the vertices denote their chiral orders.}
	\label{fig.Feynmandiagrams}
\end{figure}

Up to $\mathcal{O}(p^4)$ the decay amplitudes receive the contributions from the Feynman diagrams shown in Fig.~\ref{fig.Feynmandiagrams}. The decay amplitudes for $\eta\to\pi^+\pi^-a$ and $\eta\to\pi^0\pi^0a$ turn out to be the same up to $\mathcal{O}(p^4)$ in isospin limit. The LO part of the decay amplitude has no energy dependence and its explicit expression is 
\begin{equation}
\mathcal{M}_{\eta;\pi\pi a}^{(2)}=
\left[\frac{2}{\sqrt{3}}+\left(1-\frac{m_a^2}{m_a^2-m_{\eta}^2}\right)\frac{m_{\eta}^2-m_{\pi}^2}{\sqrt{3}m_{\eta}^2}\right]\frac{m_{\pi}^2}{6f_aF_{\pi}}\,.
\label{eq.MLO}
\end{equation}
We have replaced the LO quantities $F$, $\bar{m}_{\pi}^2$ and $\bar{m}_{\eta_8}^2$ by the physical ones in the above equation, which is legitimate at LO accuracy. While, up to NLO, this will bring an extra $\mathcal{O}(p^4)$ correction into the decay amplitude
\begin{equation}
\begin{aligned}
&\mathcal{M}_{\eta;\pi\pi a}^{(4)}\supset \mathcal{M}_{\eta;\pi\pi a}^{(2)}
\left(\delta_{F_{\pi}}^{(4)}-\frac{\delta_{m_{\pi}^2}^{(4)}}{m_{\pi}^2}\right)
+\left(1-\frac{m_{a}^2}{m_a^2-m_{\eta}^2}\right)
\left(\frac{m_{\eta}^2-m_{\pi}^2}{\sqrt{3}m_{\eta}^2}\frac{\delta_{m_{\eta}^2}^{(4)}}{m_{\eta}^2}-\frac{\delta_{m_{\eta}^2}^{(4)}-\delta_{m_{\pi}^2}^{(4)}}{\sqrt{3}m_{\eta}^2}\right)
\frac{m_{\pi}^2}{6f_aF_{\pi}}
\\ &+\frac{m_a^2}{m_a^2-m_{\eta}^2}\frac{\delta_{m_{\eta}^2}^{(4)}}{m_a^2-m_{\eta}^2}\frac{m_{\eta}^2-m_{\pi}^2}{\sqrt{3}m_{\eta}^2}\frac{m_{\pi}^2}{6f_aF_{\pi}}\,,
\end{aligned}
\label{eq.CorrectionFromLOPara}
\end{equation}
where $\delta_{F_{\pi}}^{(4)}$, $\delta_{m_{\eta}^2}^{(4)}$ and $\delta_{m_{\pi}^2}^{(4)}$ denote the $\mathcal{O}(p^4)$ corrections to the decay constant $F_\pi$, the $\eta$ mass and the pion mass, respectively. They have been worked out long time ago in the seminal paper of $SU(3)$ $\xpt$~\cite{Gasser:1984gg} and their expressions are provided here for the sake of completeness
\begin{align}
&F_{\pi}=F\left(1+\delta_{F_{\pi}}^{(4)}\right)\,,\quad
m_{\pi}^2=\bar{m}_{\pi}^2+\delta_{m_{\pi}^2}^{(4)}\,,\quad
m_{\eta}^2=\bar{m}_{\eta_8}^2+\delta_{m_{\eta}^2}^{(4)}\,,
\\
&\delta_{F_{\pi}}^{(4)}=\frac{2}{F_{\pi}^2}\left[3L_4^r(m_{\pi}^2+m_{\eta}^2)+2L_5^rm_{\pi}^2\right]
-\frac{m_{\pi}^2}{(4\pi F_{\pi})^2}\log\frac{m_{\pi}^2}{\mu^2}-\frac{m_{K}^2}{2(4\pi F_{\pi})^2}\log\frac{m_{K}^2}{\mu^2}\,,
\\
&\begin{aligned}
\delta_{m_{\pi}^2}^{(4)}=&
m_{\pi}^2\Big\{
\frac{m_{\pi}^2}{2(4\pi F_{\pi})^2}\log\frac{m_{\pi}^2}{\mu^2}
-\frac{m_{\eta}^2}{6(4\pi F_{\pi})^2}\log\frac{m_{\eta}^2}{\mu^2}
-\frac{4}{F_{\pi}^2}\big[2(L_5^r-2L_8^r)m_{\pi}^2
\\
&+3(L_4^r-2L_6^r)(m_{\pi}^2+m_{\eta}^2)
\big]
\Big\}\,,
\end{aligned}
\\
&\begin{aligned}
\delta_{m_{\eta}^2}^{(4)}=&
m_{\eta}^2\Big\{
\frac{m_{K}^2}{(4\pi F_{\pi})^2}\log\frac{m_{K}^2}{\mu^2}
-\frac{2}{3}\frac{m_{\eta}^2}{(4\pi F_{\pi})^2}\log\frac{m_{\eta}^2}{\mu^2}
-\frac{4}{F_{\pi}^2}\big[2(L_5^r-2L_8^r)m_{\eta}^2
\\
&+3(L_4^r-2L_6^r)(m_{\pi}^2+m_{\eta}^2)
\big]
\Big\}
-m_{\pi}^2\Big\{
\frac{m_{\pi}^2}{2(4\pi F_{\pi})^2}\log\frac{m_{\pi}^2}{\mu^2}
-\frac{m_{K}^2}{3(4\pi F_{\pi})^2}\log\frac{m_{K}^2}{\mu^2}
\\
&-\frac{m_{\eta}^2}{6(4\pi F_{\pi})^2}\log\frac{m_{\eta}^2}{\mu^2}
\Big\}
+\frac{8}{F_{\pi}^2}(3L_7^r+L_8^r)\left(m_{\eta}^2-m_{\pi}^2\right)^2\,.
\end{aligned}
\end{align}

The decay amplitude acquires energy dependence at NLO. As shown in Fig.~\ref{fig.Feynmandiagrams}, the NLO part of the decay amplitude receives the contributions from contact diagram in Fig.~\ref{fig.1b} by taking the $\mathcal{O}(p^4)$ vertices, and the loop diagrams in Figs.~\ref{fig.1c}-\ref{fig.1f} by taking the $\mathcal{O}(p^2)$ vertices, where the intermediate states can be $\pi\pi,\,K\bar{K},\,\eta\eta$ for the $s$-channel diagram in Fig.~\ref{fig.1d}, and $\pi\eta,\,K\bar{K}$ for the $t$- and $u$-channel diagrams in Figs.~\ref{fig.1e} and \ref{fig.1f}.
By plugging the field redefinitions of Eq.~(\ref{eq.FieldsRenormalization}) into the $\mathcal{O}(p^2)$ Lagrangian, one gets extra NLO contributions for the decay amplitude, which can be recognized as the corrections from the wave function renormalization of $\eta$ and also the NLO ALP-$\eta$ mixing. In addition, the wave function renormalization constant for the pion is also needed in the calculation, while the ALP wave function renormalization is neglected since it behaves as $Z_a=1+\mathcal{O}(1/f_a^2)$. In Appendix~\ref{appendix.A}, we give the full $\mathcal{O}(p^4)$ part of the $\eta\to\pi\pi a$ amplitude, which incorporates the correction of Eq.~(\ref{eq.CorrectionFromLOPara}). We have explicitly checked that the divergences originated from the chiral loops in the $\eta\to\pi\pi a$ amplitude can be completely absorbed by the renormalization of the $SU(3)$ $\xpt$ LECs determined in the seminal paper~\cite{Gasser:1984gg}. This also provides a way to cross check about the correctness of the decay amplitudes.

Being the momentum expansion feature of $\xpt$, its description of the $\pi\pi$ system up to the $\eta$ mass in the $\eta\to\pi\pi a$ decay could be insufficient, due to the strong $\pi\pi$ final-state interaction. The unitarization procedure offers an efficient way to extend the applicable energy range of the perturbative $\xpt$ results. While for the $a\pi$ subsystem, the $\xpt$ result up $\mathcal{O}(p^4)$, as discussed in detail later, indicates that its final-state interaction could be perturbatively treated as a good approximation, at least in the energy region up to $m_\eta-m_\pi$. 
As a result, we will focus on the unitarization of the $\pi\pi$ system in the $\eta\to\pi\pi a$ process. 
Regarding the $\eta'\to\pi\pi a$ decay, the conventional $SU(3)$ $\chi$PT is invalid and one should use the $U(3)$ $\chi$PT instead, where the $U_A(1)$ anomaly effect is explicitly included~\cite{Kaiser:2000gs,Herrera-Siklody:1996tqr} and the so-called $\delta$ counting scheme, namely $O(\delta)\sim O(p^2)\sim O(m_q)\sim O(1/N_C)$ should be employed. In the $\delta$ counting scheme, the renormalization is a bit subtle, since the LECs that renormalize the chiral loops belong to different orders. Meanwhile the chiral loops in $U(3)$ $\chi$PT belong to the next-to-next-to-leading order (NNLO) in the $\delta$ counting, and many poorly known $O(p^6)$ LECs also appear at NNLO, which will introduce uncontrolled uncertainties to the phenomenological discussions. Moreover, due to the fact $m_{\eta'}\sim$~1~GeV, the intricate massive scalar resonances $f_0(980)$ and $a_0(980)$, and the coupled $K\bar{K}$ channel effect could play visible roles in the $\eta'\to\pi\pi a$ decay. Therefore we consider it deserves another independent future work to perform the unified study of the axion production from $\eta$ and $\eta'$ decays in the $U(3)$ $\chi$PT framework. It is mentioned that a recent work in Ref.~\cite{Gao:2022xqz} has carried out the NLO calculation of the $a$-$\pi$-$\eta$-$\eta'$ mixing in $U(3)$ $\chi$PT, which provides a necessary step to study the $\eta/\eta'\to \pi\pi a$ decays.

The unitarization can be conveniently implemented through the partial-wave (PW) amplitude, which is obtained from the PW projection
\begin{equation}\label{eq.pwm}
\mathcal{M}_{\eta;\pi\pi a}^{IJ}(s)=
\frac{1}{2(\sqrt{2})^{N}}\int_{-1}^{+1}\mathrm{d}\cos\theta
P_J(\cos\theta)\mathcal{M}_{\eta;\pi\pi a}^{I}(s,t,u)\,,
\end{equation}
where $I$ and $J$ are the quantum numbers of the isospin and angular momentum of the $\pi\pi$ system respectively, $\theta$ denotes the scattering angle in the center of mass (CM) frame and $P_J(\cos\theta)$ corresponds to the Legendre polynomial. The term $(\sqrt{2})^{N}$ stands for the symmetry factor, with $N$ the number of the pairs of identical particles. $N=1$ in the $\eta\to\pi\pi a$ case. The Mandelstam variables $t$ and $u$ in the CM frame are given by 
\begin{equation}
\begin{aligned}
&t=\frac{1}{2}(m_{\eta}^2+2m_{\pi}^2+m_a^2-s)-\frac{\sqrt{\lambda(m_{\eta}^2,s,m_a^2)\lambda(s,m_{\pi}^2,m_{\pi}^2)}}{2s}\cos\theta\,,
\\
&u=\frac{1}{2}(m_{\eta}^2+2m_{\pi}^2+m_a^2-s)+\frac{\sqrt{\lambda(m_{\eta}^2,s,m_a^2)\lambda(s,m_{\pi}^2,m_{\pi}^2)}}{2s}\cos\theta\,,
\end{aligned}
\end{equation}
with $\lambda(x,y,z)=x^2+y^2+z^2-2xy-2xz-2yz$. The $P$-wave $\pi\pi$ amplitude vanishes in the $\eta\to\pi\pi a$ process. The predominant contribution comes from the $S$-wave amplitude, and the $D$-wave effect is expected to be small in the kinematical region of the $\eta\to\pi\pi a$ decay. Consequently, we will perform the unitarization of the $S$-wave $\pi\pi$ amplitude, and directly include the $D$-wave projection from the perturbative $\xpt$ results. Since we work in the isospin limit, the only possible isospin of the two pions in $\eta\to\pi\pi a$ process is $I=0$.  

Up to the $\eta$ mass, the final-state $\pi\pi$ interaction is well approximated by including the single-channel unitarity effect and we propose the following recipe~\cite{Oller:2000fj} to perform the unitarization for the $S$-wave $\pi\pi$ system
\begin{equation}\label{eq.unim}
\mathcal{M}_{\eta;\pi\pi a}^{00,{\rm Uni}}(s)=\frac{\mathcal{M}_{\eta;\pi\pi a}^{00,{\rm L}}(s)}{1-G_{\pi\pi}(s)T_{\pi\pi\to\pi\pi}^{00,(2)}(s)}\,,
\end{equation}
where the quantity $\mathcal{M}_{\eta;\pi\pi a}^{00,{\rm L}}(s)$ excludes the $s$-channel unitarity cut from the $\pi\pi$ case and contains the left hand cuts and contact terms of the perturbative $\eta\to\pi\pi a$ amplitude, $T_{\pi\pi\to\pi\pi}^{00,(2)}(s)$ stands for the LO $\pi\pi$ scattering amplitude with $IJ=00$, and $G_{\pi\pi}(s)$ is the two-point one-loop function evaluated in the dimensional regularization by introducing a subtraction constant $a_{SC}$. 
The reason that we only take the LO $\pi\pi$ amplitude to perform the unitarization, lies on the fact that the isoscalar and scalar $\pi\pi$ partial wave dominates in the $\eta\to\pi\pi a$ process. According to many previous works~\cite{Oller:1997ti,Oller:1998hw,Guo:2011pa,Guo:2012yt,Gao:2019idb}, the $\pi\pi$ phase shifts with $IJ=00$ and the $\sigma$ pole can be well reproduced when unitarizing the LO $\pi\pi$ amplitude.

The explicit expression of the $G_{\pi\pi}$ function in Eq.~\eqref{eq.unim} is 
\begin{equation}\label{eq.gfunc}
G_{\pi\pi}(s)=-\frac{1}{(4\pi)^2}\left(\log\frac{m_{\pi}^2}{\mu^2}-\sigma_{\pi}(s)\log\frac{\sigma_{\pi}(s)-1}{\sigma_{\pi}(s)+1}+a_{SC}\right)\,,
\end{equation} 
whose imaginary part is 
\begin{equation}
\mathrm{Im}G_{\pi\pi}(s)=\theta(s-4m_{\pi}^2)\rho_{\pi\pi}(s)\,,
\end{equation}
with 
\begin{equation}
\rho_{\pi\pi}(s)=\frac{\sigma_{\pi}(s)}{16\pi}\,,\quad
\sigma_{\pi}(s)=\sqrt{1-\frac{4m_{\pi}^2}{s}}\,. 
\end{equation}
The subtraction constant $a_{SC}$ introduced in the unitarization procedure will be fitted to the $\pi\pi$ phase shifts obtained from Roy equation analysis\cite{Garcia-Martin:2011iqs}. The LO PW $\pi\pi$ scattering amplitude in the convention of Eq.~\eqref{eq.pwm} with $IJ=00$ reads
\begin{equation}
T_{\pi\pi\to\pi\pi}^{00,(2)}(s)=\frac{2s-m_{\pi}^2}{2F_{\pi}^2}\,.
\end{equation}
The quantity $\mathcal{M}_{\eta;\pi\pi a}^{00,{\rm L}}(s)$ in Eq.~\eqref{eq.unim} can be determined by matching the right-hand side of this equation with the perturbative $\xpt$ result, and the explicit expression is given by 
\begin{equation}
\mathcal{M}_{\eta;\pi\pi a}^{00,{\rm L}}(s)= 
\mathcal{M}_{\eta;\pi\pi a}^{00,(2)}(s)+\mathcal{M}_{\eta;\pi\pi a}^{00,(4)}(s)-G_{\pi\pi}(s)\,\mathcal{M}_{\eta;\pi\pi a}^{00,(2)}(s)\,T_{\pi\pi\to\pi\pi}^{00,(2)}(s)\,,\label{eq.ND}
\end{equation}
which satisfies 
\begin{equation}
{\rm Im} \mathcal{M}_{\eta;\pi\pi a}^{00,{\rm L}}(s)=0\,, \qquad  (2m_\pi<\sqrt{s}<2m_K)\,,
\end{equation}
since the perturbative amplitude $\mathcal{M}_{\eta;\pi\pi a}^{00,(4)}(s)$ obeys 
\begin{equation}
{\rm Im}\mathcal{M}_{\eta;\pi\pi a}^{00,(4)}(s)= \rho_{\pi\pi}(s) \mathcal{M}_{\eta;\pi\pi a}^{00,(2)}(s)\,T_{\pi\pi\to\pi\pi}^{00,(2)}(s)\,,  \qquad  (2m_\pi<\sqrt{s}<2m_K)\,. 
\end{equation}
This matching procedure of the unitarized amplitude with the perturbative one also helps to eliminate the possible double-counting terms in the construction of unitarized expression.

It is easy to verify that the unitarized decay amplitude in Eq.~\eqref{eq.unim} strictly fulfills the unitarity relation
\begin{equation}
{\rm Im}\mathcal{M}_{\eta;\pi\pi a}^{00,{\rm Uni}}(s) = \rho_{\pi\pi} \mathcal{M}_{\eta;\pi\pi a}^{00,{\rm Uni}}(s) \left(T_{\pi\pi\to\pi\pi}^{00,{\rm Uni}}(s)\right)^*\,, \qquad  (2m_\pi<\sqrt{s}<2m_K)\,,
\end{equation}
when taking the unitarized $\pi\pi$ amplitude as 
\begin{equation}\label{eq.unit}
T_{\pi\pi\to\pi\pi}^{00,{\rm Uni}}(s)=\frac{T_{\pi\pi\to\pi\pi}^{00,(2)}(s)}{1-G_{\pi\pi}(s)T_{\pi\pi\to\pi\pi}^{00,(2)}(s)}\,,
\end{equation}
which is a generalized $K$-matrix amplitude. The common quantity of $1/[1-G_{\pi\pi}(s)T_{\pi\pi\to\pi\pi}^{00,(2)}(s)]$ in Eqs.~\eqref{eq.unim} and \eqref{eq.unit} practically plays the role of parameterizing the strong $\pi\pi$ final-state interactions. It will be demonstrated later that the unitarized $\pi\pi$ amplitude can describe the $\pi\pi$ phase shifts with $IJ=00$ very well up to $m_\eta$, implying that the neat forms in Eqs.~\eqref{eq.unim} and \eqref{eq.unit} can capture the key the unitarization effects in the $\pi\pi$ channel.  

For the $D$-wave component, we will simply take the result from the PW projection of the perturbative $\xpt$ amplitude in Eq.~\eqref{eq.pwm}. To combine the unitarized $S$-wave amplitude and the perturbative $D$-wave amplitude, the $\eta\to\pi\pi a$ decay amplitude can be reconstructed as 
\begin{equation}\label{eq.cpw}
\mathcal{M}_{\eta;\pi^{+}\pi^{-} a}^{I=0,{\rm CPW}}(s,t,u)=
\sqrt{2}\left[\mathcal{M}_{\eta;\pi\pi a}^{00,{\rm Uni}}(s)+5P_2(\cos\theta)\mathcal{M}_{\eta;\pi\pi a}^{02}(s)\right]\,,
\end{equation}
where the factor $\sqrt{2}$ is introduced to recover the physical amplitude from the PW convention in Eq.~\eqref{eq.pwm}. 
From the difference of the perturbative full amplitude $\mathcal{M}_{\eta;\pi\pi a}^{I=0}(s,t,u)$ given in the Appendix and the combined partial-wave (CPW) decay amplitude $\mathcal{M}_{\eta;\pi\pi a}^{I=0,{\rm CPW}}(s,t,u)$ in Eq.~\eqref{eq.cpw}, one could study how the strong final-state $\pi\pi$ interactions can affect the $\eta\to\pi\pi a$ process.

\section{Phenomenological discussions}\label{sec.pheno}

As mentioned previously, the isospin breaking effects enter the $\eta\to\pi\pi a$ process at the level of $\mathcal{O}(\epsilon_I^2)$ and hence are neglected in this work. In the isospin limit, the decay amplitudes for $\eta\to\pi^+\pi^-a$ and $\eta\to\pi^0\pi^0a$ turn out to be the same up to NLO. The decay rate for the $\pi^0\pi^0$ final state only differs with that of $\pi^+\pi^-$ by an identical-particle factor $1/2$. So we only show the results for the $\eta\to\pi^+\pi^-a$ decay in the following. 

In the phenomenological discussions, we take $m_{\pi}=137~$MeV, $m_{K}=496~$MeV, and $m_{\eta}=548~$MeV. The value of the pion decay constant is taken as $F_{\pi}=92.1~$MeV~\cite{ParticleDataGroup:2022pth}. Regarding the renormalized LECs $L_i^{r}(\mu)$ at the renormalization scale $\mu=770~$MeV, we take the values fitted at NLO given in Table~6 of Ref.~\cite{Bijnens:2014lea}, which are $L_1^r=1.0(1)\times 10^{-3}$, $L_2^r=1.6(2)\times 10^{-3}$, $L_3^r=-3.8(3)\times 10^{-3}$, $L_4^r=0.0(3)\times 10^{-3}$, $L_5^r=1.2(1)\times 10^{-3}$, $L_6^r=0.0(4)\times 10^{-3}$, $L_7^r=-0.3(2)\times 10^{-3}$ and $L_8^r=0.5(2)\times 10^{-3}$.  
To estimate the uncertainty of the $\eta\to\pi\pi a$ amplitude, we will generate a large amount of parameter configurations by randomly sampling all the $L_{i=1,\cdots,8}^r$ with the assumptions that they are Gaussianly distributed and not correlated with each other. Since the latter assumption clearly introduces unrealistically large uncertainty, in another scenario we will simply neglect the uncertainties of the LECs of $L_{4,6,7}^r$ and only consider the error bars of $L_{1,2,3,5,8}^r$ in the random sampling procedure. In later discussions, the uncertainty analysis by taking the error bars of all the $L_{i=1,\cdots,8}^r$ in the random sampling  will be named as Set-I, while the uncertainty analysis by keeping the error bars of $L_{1,2,3,5,8}^r$ in the random sampling procedure will be denoted as Set-II.

The differential partial decay rate for the $\eta(P_{\eta}) \to \pi(p_1)\pi(p_2)a(p_a)$ process reads
\begin{equation}
\frac{\mathrm{d}^2\Gamma_{\eta;\pi\pi{a}}}{\mathrm{d}s\mathrm{d}t}=\frac{1}{32(2\pi)^3\mathcal{N}m_{\eta}^3}
\left|\mathcal{M}_{\eta;\pi\pi{a}}\right|^2\,,\label{eq.DifferentialWidth}
\end{equation}
with $\mathcal{N}=1$ for $\pi^+\pi^-$ channel and $\mathcal{N}=2$ for $\pi^0\pi^0$ channel. The boundaries of the physical regions for $s$ and $t$ are
\begin{align}
&s_{\text{min}}=4m_{\pi}^2\,,\quad s_{\text{max}}=(m_{\eta}-m_a)^2\,,
\\
&t_{\text{max/min}}(s)=\frac{1}{2}\Bigg[m_{\eta}^2+m_a^2+2m_{\pi}^2-s\pm
\frac{\sqrt{\lambda(s,m_{\eta}^2,m_a^2)\lambda(s,m_{\pi}^2,m_{\pi}^2)}}{s}
\Bigg]\,,
\end{align}
or 
\begin{align}
&t_{\mathrm{min}}=(m_{\pi}+m_a)^2\,,\quad t_{\mathrm{max}}=(m_{\eta}-m_{\pi})^2\,,
\\
&\begin{aligned}
s_{\mathrm{max/min}}(t)=& 
\frac{m_{\eta}^2+m_a^2+2m_{\pi}^2-t}{2}+\frac{1}{2t}(m_{\eta}^2-m_{\pi}^2)(m_{\pi}^2-m_a^2)
\pm\frac{\sqrt{\lambda(t,m_{\eta}^2,m_{\pi}^2)\lambda(t,m_{\pi}^2,m_a^2)}}{2t}\,.
\end{aligned}
\end{align}

\begin{figure}[htbp]
\centering
\includegraphics[width=0.95\textwidth]{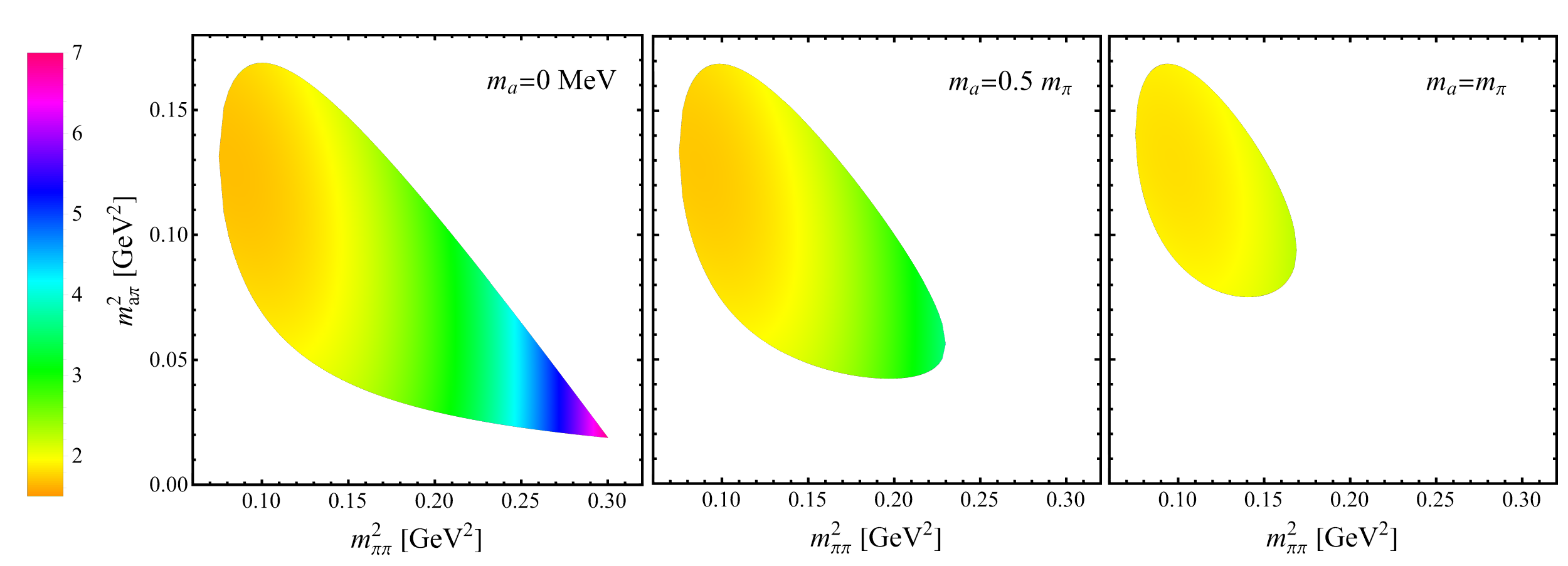}
\caption{$10^6\times f_a^2\mathrm{d}^2\Gamma/\mathrm{d}s\mathrm{d}t$ (see Eq.~\eqref{eq.DifferentialWidth}) in unit of $\mathrm{GeV}^{-1}$. The results correspond to taking all the three terms in the right hand side of Eq.~\eqref{eq.m2exp}, i.e. the $\mathrm{NLO'}$ case. See the text for details. \label{fig.dalitz_NLO2}}
\end{figure}

\subsection{Analyses by taking the perturbative $\chi$PT amplitudes}

In this part, we focus on the situation by taking the full perturbative $\chi$PT amplitudes $\mathcal{M}_{\eta;\pi\pi a}^{(2+4)}(s,t,u)$, as given in Eq.~\eqref{eq.decayampfull}. 
When perturbatively including the NLO corrections in the decay amplitude, we can write the amplitude squared as
\begin{equation}\label{eq.m2exp}
|\mathcal{M}|^2=\left(\mathcal{M}^{(2)}\right)^2+2\mathcal{M}^{(2)}\mathrm{Re}\left(\mathcal{M}^{(4)}\right)+|\mathcal{M}^{(4)}|^2\,,
\end{equation}
where $n$ inside the bracket in the superscript stands for the chiral order of the amplitude. The well established power counting in $\xpt$ is analyzed in most of the cases for the amplitude itself, instead of the decay width that takes the amplitude squared as input. By applying the chiral power counting to the amplitude squared, the $|\mathcal{M}^{(4)}|^2$ term in Eq.~\eqref{eq.m2exp} belongs to the $O(p^6)$ part, which would have the same order as $\mathcal{M}^{(2)}\mathrm{Re}\left(\mathcal{M}^{(6)}\right)$. The latter term requires the calculation of the next-to-next-to-leading order amplitude, which is beyond the scope of the present study. However, by applying the chiral power counting rule to the amplitude itself, one should take all the three terms in Eq.~\eqref{eq.m2exp}, which guarantees the positivity of the amplitude squared by construction. In this work, we will explore the consequences of the two scenarios by taking the chiral power counting rule both to the amplitude itself and the amplitude squared. The difference from the two scenarios can be regarded as a measure of the higher order effects. 
We designate the decay rate by keeping the first two terms in the amplitude squared in Eq.~\eqref{eq.m2exp} as $\Gamma_{\mathrm{NLO}}$, and denote $\Gamma_{\mathrm{NLO'}}$ when taking all the three terms in the previous equation. The Dalitz decay distributions from the $\mathrm{NLO'}$ scenario are shown in Fig.~\ref{fig.dalitz_NLO2}, where we take three values of $m_a$: $m_a=0~$MeV, $m_a=0.5m_{\pi}$, $m_a=m_{\pi}$, for illustrations. The first lesson we can learn is that the larger the ALP mass the smoother of the Dalitz distributions result. The $\eta\to\pi\pi a$ amplitude up to NLO becomes roughly flat in the $m_{\pi\pi}^2$-$m_{a\pi}^2$ plane for $m_a\gtrsim m_\pi$. Another important lesson is that almost for any given value of $m_{\pi\pi}^2$ the distribution along the $m_{a\pi}^2$ direction is rather smooth, while in contrast for a specific $m_{a\pi}^2$ value the distribution along the $m_{\pi\pi}^2$ direction can be uneven. 
This gives hints that the $a\pi$ interaction strength is weak in the kinematical region allowed in the $\eta\to\pi\pi a$ process and the $\pi\pi$ final-state interaction is stronger, which validates the discussions in the previous section that focuses on the $\pi\pi$ final-state interactions. It is also possible that in a broader energy region there are resonant contributions, such as $a_0(980)$ and $a_0(1450)$, to the $a\pi$ channel, which are however distant from the kinematical region of the $\eta\to\pi\pi a$ decay. This latter scenario could also give an even distribution along the $m_{a\pi}^2$ direction in the $\eta\to\pi\pi a$ decay.

\begin{figure}[htbp]
\centering
\includegraphics[width=0.95\textwidth]{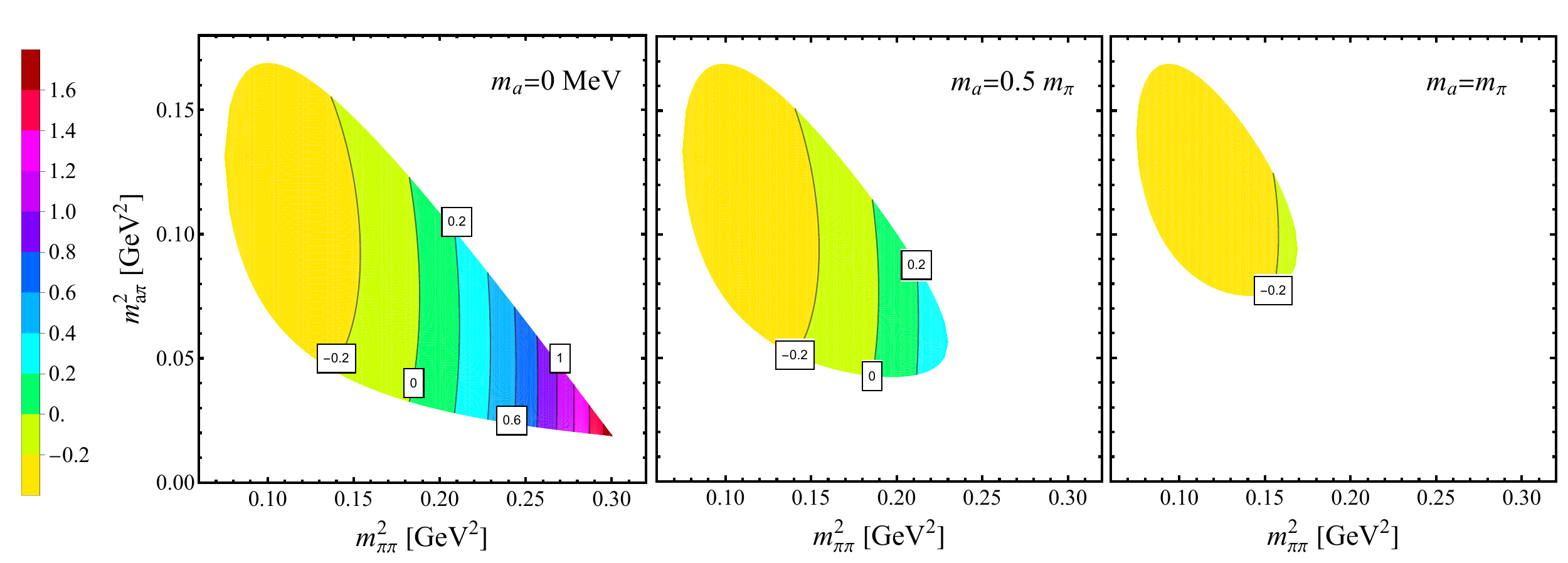}
\caption{The ratios of $|\mathcal{M}|_{\rm NLO'}^2/|\mathcal{M}|_{\rm LO}^2$, where $|\mathcal{M}|_{\rm LO}^2$ and $|\mathcal{M}|_{\rm NLO'}^2$ stand for the first and last two terms in Eq.~\eqref{eq.m2exp}, respectively.  }\label{fig.dalitz_nlo2vslo}
\end{figure}

The ratios between the higher order contribution and the LO part of the amplitude squared~\eqref{eq.m2exp} in the $m_{\pi\pi}^2$-$m_{a\pi}^2$ plane are visualized in Fig.~\ref{fig.dalitz_nlo2vslo}, from which one can have an intuitive idea about the convergence of the chiral expansion of the amplitude squared $|\mathcal{M}|^2$. The fact that the contour lines are roughly parallel to the $m_{a\pi}^2$ axis semi-quantitatively illustrates the even distributions of the amplitude squared along the $m_{a\pi}^2$ direction. For $m_a\simeq m_\pi$, the $\mathrm{NLO'}$ correction to the amplitude squared, with respect to the LO result, lies around at the level of $20\%\sim 40\%$ in the whole $m_{\pi\pi}^2$-$m_{a\pi}^2$ plane. In contrast, for the QCD axion case with $m_a\sim 0$, the $\mathrm{NLO'}$ correction tends to rapidly increase when $m_{\pi\pi} \gtrsim 0.48$~GeV. In the small area for $0.5~{\rm GeV}< m_{\pi\pi} < m_{\eta}$, the ratios between the $\mathrm{NLO'}$ correction and the LO result can be 100\% and even larger.

\begin{figure}[htbp]
\centering
\includegraphics[width=\textwidth]{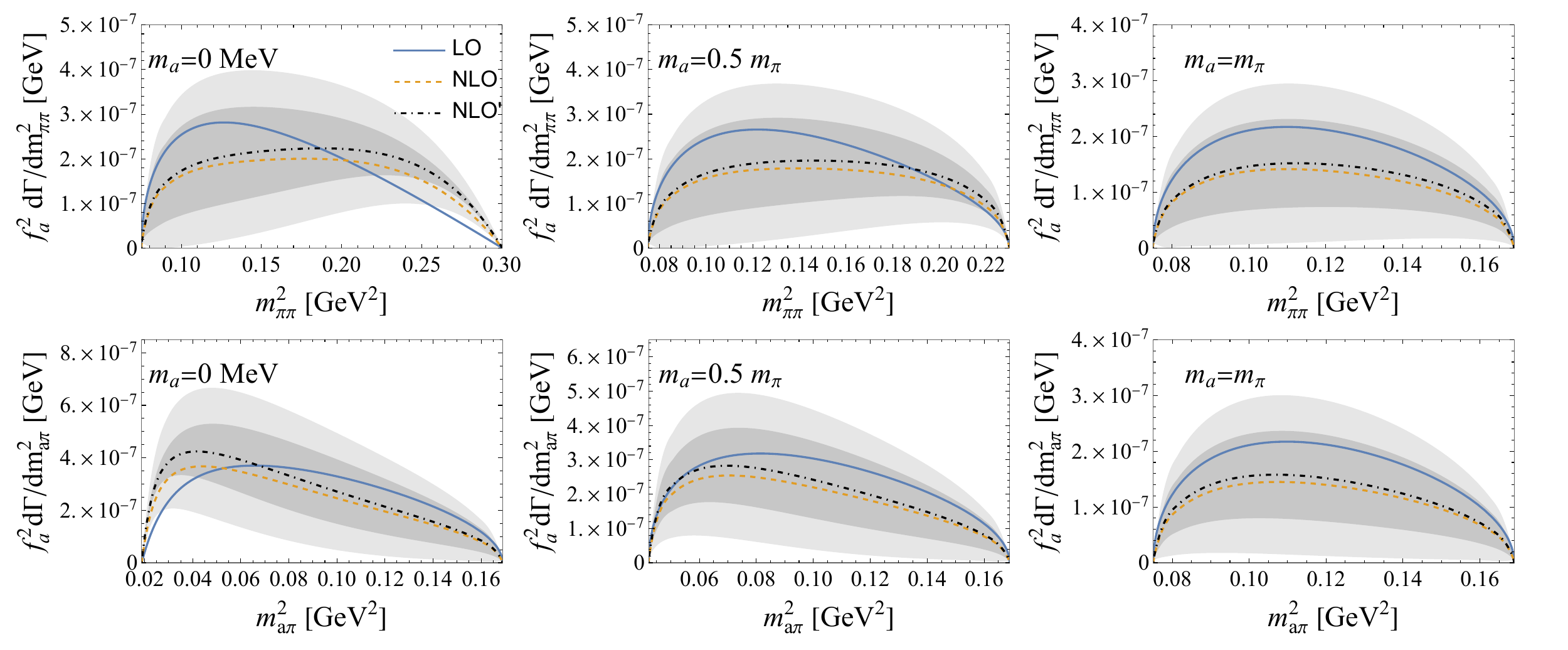}
\caption{The invariant-mass spectra of the $\pi\pi$ (upper row) and $a\pi$ (lower row) systems in the $\eta\to\pi\pi a$ process.  The spectra are shown by taking three different values of $m_a$, as explicitly indicated in the plots. In each panel, we show three different curves by taking the amplitudes squared in the LO, NLO, and NLO$'$ cases. The lighter and darker shaded regions correspond to the error estimations for the NLO$'$ results from the Set-I and Set-II scenarios, respectively. See the text for further explanations. }
\label{fig.twobodyspect}
\end{figure}

\begin{figure}[htbp]
\centering
\includegraphics[width=0.99\textwidth]{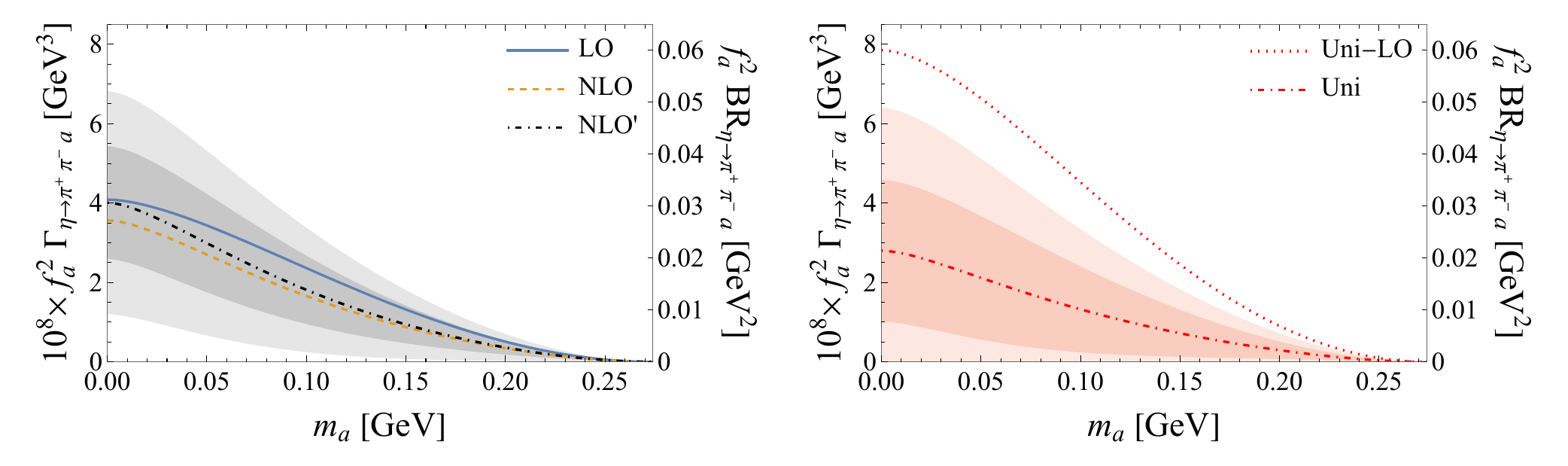}
\caption{The partial decay widths (left axis) and branching ratios (right axis) at different values of $m_a$ for the $\eta\to\pi^+\pi^-a$ process. The curves labeled as LO, NLO and NLO$'$ (shown in the left panel) are obtained by taking the perturbative $\xpt$ amplitudes, while the curves labeled as Uni-LO and Uni (shown in the right panel) correspond to the results by taking the unitarized amplitudes in Eqs.~\eqref{eq.unimlo} and \eqref{eq.unim}, respectively. See the text for the details of different notations.}
\label{fig.widthma}
\end{figure}

By integrating over the $t$ or $s$ in the Dalitz distribution via  Eq.~(\ref{eq.DifferentialWidth}), one can obtain the $\pi\pi$ or $a\pi$ spectra in the $\eta\to\pi^+\pi^-a$ decay, and the results from the LO, NLO and $\mathrm{NLO'}$ cases are summarized together in Fig.~\ref{fig.twobodyspect} with different values of $m_a$. 
By integrating over both $s$ and $t$ variables in Eq.~\eqref{eq.DifferentialWidth}, one gets the partial decay width of $\eta\to\pi^+\pi^-a$ and the results with varying $m_a$ are shown in Fig.~\ref{fig.widthma}. Large uncertainties represented by the shaded areas are observed in Figs.~\ref{fig.twobodyspect} and \ref{fig.widthma}. In Appendix~\ref{appendix.A}, it is demonstrated that by using the lowest resonance saturation advocated in Ref.~\cite{Ecker:1988te} the parts of the $\mathcal{O}(p^4) $ LECs in the $\eta\to\pi\pi a$ decay are found to be mainly saturated by the scalar resonances at large $N_C$. Phenomenologically speaking, the couplings and the masses of the scalar resonances in the large $N_C$ case are still under debate~\cite{Guo:2021blc}, and this  provides an alternative way to understand the large uncertainties seen in the $\eta\to\pi\pi a$ amplitudes at $\mathcal{O}(p^4)$.  

According to the curves in Fig.~\ref{fig.widthma}, the partial decay widths of the $\eta\to\pi^+\pi^-a$ process calculated from the amplitudes squared from the LO, NLO and $\rm{NLO'}$ cases look roughly similar when varying $m_a$ from 0 to $m_{\eta}-2m_{\pi}$. 
It is verified that there is an accidental cancellation among different pieces at $O(p^4)$, i.e., the contributions from the chiral loops and the $O(p^4)$ LECs with the central values are cancelled to some extent in the $\eta\to\pi\pi a$ amplitude, which explains the similar curves obtained from the LO, NLO and NLO' cases. By taking a closer look at the upper left panel of Fig.~\ref{fig.twobodyspect} with $m_a=0$, one could gain further insights into the cancellation: the $O(p^4)$ corrections tend to decrease/increase the LO results in the lower/larger $m_{\pi\pi}$ region, which makes the integration of the three kinds of amplitudes over the entire $m_{\pi\pi}$ range similar. However, when taking somewhat different values of the $O(p^4)$ LECs from their central ones, there is no obvious cancellation and the correction from the $O(p^4)$ part becomes more important. So that the resulting curves at $O(p^4)$ can clearly deviate from the LO one, see the boundaries of the error bands shown in Figs.~\ref{fig.twobodyspect} and \ref{fig.widthma}.
In fact, the differences resulting by taking the three types of amplitudes squared when using the central values of the $O(p^4)$ LECs are obviously smaller than the large uncertainties caused by the error bars of the $\mathcal{O}(p^4)$ LECs. Furthermore, we have separately examined the uncertainty of the $\eta\to\pi\pi a$ amplitude caused by each LEC and found that the error bars of $L_{i=2,7}$ play the most important roles in the uncertainty analysis of the $\eta\to\pi\pi a$ amplitude. Therefore we point out that to further pin down their error bars will be a future step to reach a more precise prediction to the $\eta\to\pi\pi a$ decay. 

Before ending this subsection, we would like to clarify the subtle issue when the ALP mass approaches the pion mass in the predictions to the partial decay widths shown in Fig.~\ref{fig.widthma}. Such predictions around the $m_a\simeq m_\pi$ regions in previous works~\cite{Gan:2020aco,Alves:2020xhf} are usually shaded out, due to the pole singularity introduced via the ALP-$\pi^0$ mixing formula, which looks like $\theta_{a\pi}=\delta_{a\pi}/(m_\pi^2-m_a^2)$, being $\delta_{a\pi}$ the $a$-$\pi$ mixing strength. It should be mentioned that the latter mixing formula is typically obtained under the weak mixing assumption, i.e., $|\theta_{a\pi}|\ll 1$. However, in the vicinity of $m_a=m_\pi$, the approximated $\theta_{a\pi}=\delta_{a\pi}/(m_\pi^2-m_a^2)$ mixing formula does not hold any more. In fact, one does not expect any pole singularity from the mixing. In our calculation, we have taken the isospin limit, under which the ALP-$\pi^0$ mixing exactly vanishes. This naturally avoids the inclusion of the subtle ALP-$\pi^0$ mixing formula in the vicinity of $m_a=m_\pi$. Since the ALP-$\pi^0$ mixing contributes at the order of $\epsilon_I^2$ and there is no pole singularity, it is reasonable to assume that the ALP-$\pi^0$ mixing only gives smooth and mild effects. Therefore we consider our treatment by first taking the isospin limit provides a way to give reliable predictions in the vicinity of $m_a=m_{\pi}$ for the $\eta\to\pi\pi a$ partial decay widths shown in Fig.~\ref{fig.widthma}.

\subsection{Results by including the $\pi\pi$ final-state interaction}

In this part, we analyze the results after including the $\pi\pi$ final-state interactions via the unitarization procedure explained in Sec.~\ref{sec.decayamp}. The amplitude squared $|\mathcal{M}^{\rm CPW}|^2$ will be evaluated by taking the combined partial-wave amplitude $\mathcal{M}_{\eta;\pi\pi a}^{I=0,{\rm CPW}}$ given in Eq.~\eqref{eq.cpw}. In this case, the energy dependence of the $a\pi$ amplitude is totally encoded in the $D$-wave part, i.e. the second term of Eq.~\eqref{eq.cpw}. 

\begin{figure}[htbp]
\centering
\includegraphics[width=0.8\textwidth]{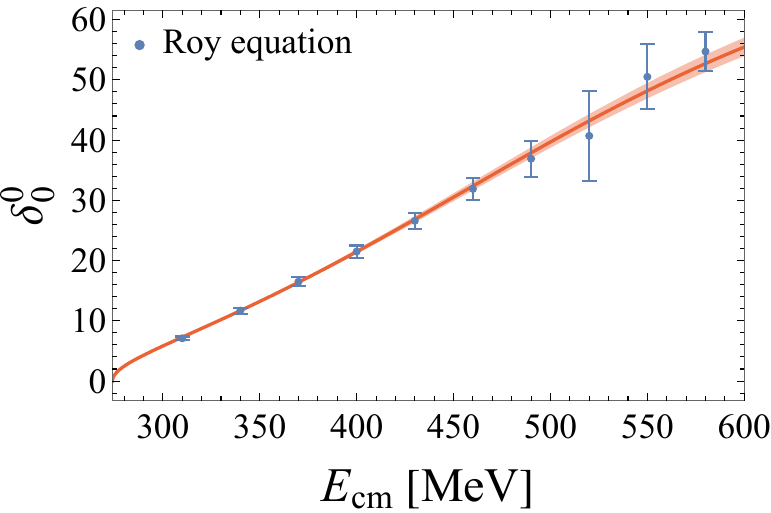}
\caption{ Comparison between the $\pi\pi$ phases in the unitarized $\eta\to\pi\pi a$ amplitude~\eqref{eq.unim} and the $\pi\pi$ scattering phase shifts from Roy equation analyses~\cite{Garcia-Martin:2011iqs}. The shaded area corresponds to the uncertainty at 1-$\sigma$ level, which is obtained by varying the parameter $a_{SC}$ in the range of Eq.~\eqref{eq.ascnum}. }
\label{fig.deltapipi00}
\end{figure}

By construction, the phases of the unitarized $S$-wave decay amplitude in Eq.~\eqref{eq.unim} equal to the $\pi\pi$ scattering phase shifts given by Eq.~\eqref{eq.unit}.  To fit the $\pi\pi$ phase shifts in the $IJ=00$ channel from the rigorous Roy equation analyses~\cite{Garcia-Martin:2011iqs}, the subtraction constant $a_{SC}$ introduced in the unitarization procedure in Eq.~\eqref{eq.gfunc} is determined to be
\begin{equation}\label{eq.ascnum}
a_{SC}=-0.97\pm 0.11 \,.
\end{equation} 
It is verified that the uncertainties of the $\eta\to\pi\pi a$ partial decay widths inherited from the error bar of $a_{SC}$ are negligible comparing with those from the $O(p^4)$ LECs. The excellent reproduction of the $\pi\pi$ phase shifts is shown in Fig.~\ref{fig.deltapipi00}. The remarkable agreements of the two results justify the use of the neat formalism in Eq.~\eqref{eq.unim} for the unitarization of the $\eta\to\pi\pi a$ decay amplitude. 

\begin{figure}[htbp]
\centering
\includegraphics[width=0.8\textwidth]{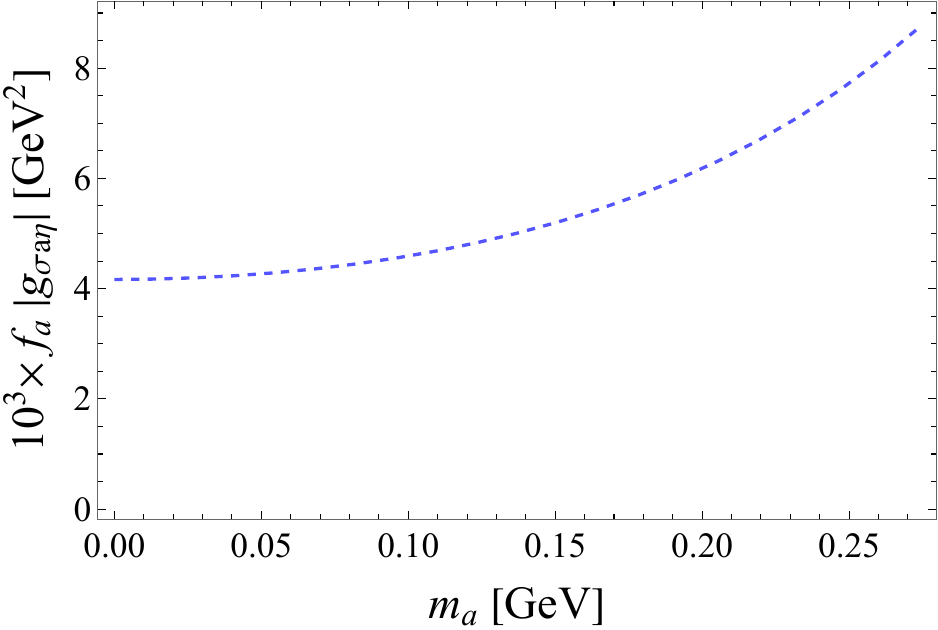}
\caption{ The axion-$\eta$-$\sigma$ coupling $|g_{\sigma a \pi}|$ multiplied by $f_a$ as a function of the ALP mass $m_a$.  }
\label{fig.gf0aeta}
\end{figure}

After the unitarization procedure, the $f_0(500)$ or $\sigma$ resonance is generated in the $\eta\to\pi\pi a$ decay amplitude and the $\pi\pi\to\pi\pi$ scattering amplitude. The resonance pole position can be calculated by extrapolating the amplitudes into the second Riemann sheet (RS). The expressions in Eqs.~\eqref{eq.unim} and \eqref{eq.unit} denote the amplitudes on the first/physical RS. The correspondirealng amplitudes on the second RS can be obtained by replacing the $G_{\pi\pi}(s)$ function with its expression on the second RS 
$G_{\pi\pi}^{\rm II}(s)$, whose explicit formula is 
\begin{equation}
G_{\pi\pi}^{\rm II}(s)= G_{\pi\pi}(s) - 2i\rho_{\pi\pi}(s)\,. 
\end{equation}
To replace the $G_{\pi\pi}(s)$ with its counter part $G_{\pi\pi}^{\rm II}(s)$ in the unitarized amplitudes in Eqs.~\eqref{eq.unim} and \eqref{eq.unit}, one can calculate the pole position and its residues in the complex energy plane. The $\sigma$ pole position is found to be $457-i253$~MeV, which is close to the rigorous dispersive analysis~\cite{Pelaez:2021dak}. Around the $\sigma$ resonance pole, the unitarized $\eta\to\pi\pi a$ amplitude on the second RS can be written as 
\begin{equation}
\mathcal{M}_{\eta;\pi\pi a}^{00,{\rm Uni, II}}(s)\big|_{s\to s_{\sigma}}\sim-\frac{g_{\sigma\pi\pi}g_{\sigma a\eta}}{s-s_{\sigma}}\,, 
\end{equation}
where the product of the couplings $g_{\sigma\pi\pi}g_{\sigma a\eta}$ corresponds to the residue. Furthermore, the $g_{\sigma\pi\pi}$ coupling can be determined from the unitarized $\pi\pi\to\pi\pi$ scattering amplitude on the second RS. Therefore, we can predict the axion-$\eta$-$\sigma$ coupling $|g_{\sigma a\eta}|$ and the results are shown in Fig.~\ref{fig.gf0aeta}. It is clear that the value of $|g_{\sigma a\eta}|$ increases when increasing the ALP masses $m_a$. 

Apart from the unitarized formula in Eq.~\eqref{eq.unim}, we also explore another scenario by only including the LO $\eta\to\pi\pi a$ amplitude in the numerator of the former equation, i.e., 
\begin{equation}\label{eq.unimlo}
\mathcal{M}_{\eta;\pi\pi a}^{00,{\rm Uni-LO}}(s)= \frac{\mathcal{M}_{\eta;\pi\pi a}^{00,(2)}(s)}{1-G_{\pi\pi}(s)T_{\pi\pi\to\pi\pi}^{00,(2)}(s)}\,,
\end{equation}
which resembles the setup employed in Ref.~\cite{Alves:2024dpa} that also uses the LO $\eta\to\pi\pi a$ amplitude in the construction of final results. The difference is that the Omn\`es function is used to take care of the final-state $\pi\pi$ interaction in the former reference. The good reproduction of the $\pi\pi$ scattering phase shifts by the unitarization factor $1/[1-G_{\pi\pi}(s)T_{\pi\pi\to\pi\pi}^{00,(2)}(s)]$, as shown in Fig.~\ref{fig.deltapipi00}, indicates that the $\pi\pi$ final-state interactions are properly included in our unitarized decay amplitudes. Furthermore, when including the higher order effects in the $\eta\to\pi\pi a$ amplitudes, such as those in Eq.~\eqref{eq.ND}, one can order by order match the unitarized decay amplitudes to the perturbative $\xpt$ results, which is a merit of our unitarized method. 

The partial decay widths calculated by taking the two types of unitarized amplitudes in Eqs.~\eqref{eq.unim} and \eqref{eq.unimlo}, together with the perturbative $\xpt$ results, are shown in Fig.~\ref{fig.widthma}. 
It is interesting to point out that the conclusion of Ref.~\cite{Alves:2024dpa} is confirmed in our study: when only taking the LO $\eta\to\pi\pi a$ amplitude as input to proceed the unitarization, the unitarized decay amplitude will lead to a clear enhancement around two times compared to the LO result. However, according to the curves in Fig.~\ref{fig.widthma}, our study shows that when including the NLO $\eta\to\pi\pi a$ amplitude in the unitarization procedure the resulting partial decay widths are not necessarily enhanced, compared to the LO result.    
In fact, when taking the central value of the LECs in Ref.~\cite{Bijnens:2014lea}, the partial decay widths from the unitarized amplitudes are even decreased. Nevertheless, we have explicitly verified that the uncertainties from the unitarized amplitudes that are caused by the LECs are also quite large, and the error bands are qualitatively similar to the shaded areas shown in Fig.~\ref{fig.widthma} from the perturbative $\xpt$ analyses.

\section{Summary and Conclusions}\label{sec.concl}

In this work, we have performed the complete calculation of the $\eta\to\pi\pi a$ amplitude up to one-loop level within the $SU(3)$ axion chiral perturbation theory in the isospin limit. A bare axion mass term is kept in the calculation, which allows us to explore the phenomenological consequences by varying the ALP mass from zero to $m_\eta-2m_\pi$. As a byproduct, we have also worked out the ALP-$\eta$ mixing and the ALP mass up to the next-to-leading order. It is verified that the $\eta\to\pi\pi a$ amplitude can be renormalized with the conventional renormalization conditions of the $SU(3)$ low energy constants. The unitarized $\eta\to\pi\pi a$ decay amplitude is further constructed to implement the $\pi\pi$ final-state interaction.

We conduct the phenomenological discussions of the $\eta\to\pi\pi a$ decay, by taking the recent determinations of the $SU(3)$ $\mathcal{O}(p^4)$ low energy constants from Ref.~\cite{Bijnens:2014lea}. We give the separate analyses by taking the perturbative $\xpt$ amplitudes and the unitarized ones with $\pi\pi$ final-state interaction. The Dalitz analyses from the perturbative amplitudes reveal that the distribution along the $a\pi$ energy is rather smooth, while the distribution along the $\pi\pi$ energy is uneven. For small values of $m_a$, the next-to-leading order correction with respect to the leading-order result can be larger than 60\% in the small region with $m_{\pi\pi} \gtrsim 0.5$~GeV. The perturbative analyses of the Dalitz distributions indicate that the $\pi\pi$ final-state interactions could play some relevant roles in the $\eta\to\pi\pi a$ decay, while the $a\pi$ interaction could be treated perturbatively at least in the kinematical region of the $\eta\to\pi\pi a$ process. The unitarized $\eta\to\pi\pi a$ decay amplitudes are constructed under these two requirements, i.e., by nonperturbatively including the $\pi\pi$ strong final-state interaction and perturbatively including the weak $a\pi$ interaction. 
The $\pi\pi$ phases in the unitarized $\eta\to\pi\pi a$ decay amplitude are consistent with the $\pi\pi$ scattering phase shifts from the rigorous Roy equation study. 
The $\eta\to\pi\pi a$ decay widths are predicted as a function of $m_a$ in different scenarios by including the leading-order, the next-to-leading order perturbative $\xpt$ amplitudes, and also the two different unitarized amplitudes. We find that the unitarized decay amplitude by only taking the leading-order chiral amplitude as input indeed can obviously enhance the $\eta\to\pi\pi a$ decay width, as observed in Ref.~\cite{Alves:2024dpa}. However, the more sophisticated unitarized amplitude by including the next-to-leading order perturbative $\xpt$ expressions does not always enhance the decay width, depending on the input values of the low energy constants. The uncertainties in the $\eta\to\pi\pi a$ decay caused by nowadays determinations of the $SU(3)$ low energy constants turn out to be quite large. To further pin down the error bars of such low energy constants will be very helpful to give more precise predictions to the $\eta\to\pi\pi a$ decay process. 

This work gives a thorough theoretical calculation of the $\eta\to\pi\pi a$ process, which offers an important channel to search the ALP at various current and future experiments in the mass range from zero to $m_\eta-2m_{\pi}$. The ALP produced in $\eta\to\pi\pi a$ could further decay into $\gamma\gamma$, $e^+e^-$, $\mu^+\mu^-$, which are subject to the model dependent axion interactions. The thoroughgoing study of the experimental signatures clearly deserves an independent future work.

\section*{Acknowledgements}

We would like to thank Sergi Gonz\`alez-Sol\'is for communications. 
This work is funded in part by the National Natural Science Foundation of China (NSFC) under Grants Nos.~12150013, 11975090, and the Science Foundation of Hebei Normal University with contract No.~L2023B09.  

\appendix

\section{Full $\mathcal{O}(p^4)$ amplitude for the $\eta\to\pi\pi a$ decay}\label{appendix.A}

The calculation of the amplitudes is carried out with the help from the computational tools of FeynCalc~\cite{Mertig:1990an,Shtabovenko:2016sxi,Shtabovenko:2020gxv} and Package-X~\cite{Patel:2015tea}. 
The explicit expressions of the relevant two-point 1PI amplitudes to this work are 

\begin{equation}
\begin{aligned}
\Sigma_{\pi\pi}^{(4)}(p^2)=&-\frac{12 L_4 p^2 \left(m_{\eta}^2+m_{\pi}^2\right)}{F_{\pi}^2}-\frac{8 L_5 m_{\pi}^2 p^2}{F_{\pi}^2}+\frac{24 L_6 m_{\pi}^2 \left(m_{\eta}^2+m_{\pi}^2\right)}{F_{\pi}^2}+\frac{16 L_8 m_{\pi}^4}{F_{\pi}^2}
\\
&+\frac{m_{\pi}^2  A_0(m_{\eta})}{96 \pi ^2 F_{\pi}^2}
+\frac{\left(3 m_{\eta}^2-4 m_K^2+5 m_{\pi}^2-4 p^2\right)A_0(m_K)}{192 \pi ^2 F_{\pi}^2 }
+\frac{ \left(m_{\pi}^2-4 p^2\right) A_0(m_{\pi})}{96 \pi ^2 F_{\pi}^2 }\,,
\end{aligned}
\end{equation}
\begin{equation}
\begin{aligned}\label{eq.Sigma_etaeta}
\Sigma_{\eta\eta}^{(4)}(p^2)=
&-\frac{12 L_4 p^2 \left(m_{\eta}^2+m_{\pi}^2\right)}{F_{\pi}^2}-\frac{8 L_5 m_{\eta}^2 p^2}{F_{\pi}^2}+\frac{24 L_6 m_{\eta}^2 \left(m_{\eta}^2+m_{\pi}^2\right)}{F_{\pi}^2}
\\
&+\frac{24 L_7 \left(m_{\eta}^2-m_{\pi}^2\right)^2}{F_{\pi}^2}
+\frac{8 L_8 \left(3 m_{\eta}^4-2 m_{\eta}^2 m_{\pi}^2+m_{\pi}^4\right)}{F_{\pi}^2}
+\frac{m_{\pi}^2 A_0(m_{\pi})}{32 \pi ^2 F_{\pi}^2}
\\
&-\frac{\left(-9 m_{\eta}^2+12 m_K^2+m_{\pi}^2+12 p^2\right)A_0(m_K)}{192 \pi ^2 F_{\pi}^2}
+\frac{\left(4 m_{\eta}^2-m_{\pi}^2\right) A_0(m_{\eta})}{96 \pi ^2 F_{\pi}^2 }\,,
\end{aligned}
\end{equation}

\begin{equation}
\begin{aligned}
\Sigma_{a\eta}^{(4)}(p^2)=
&-\frac{2 \sqrt{3} L_4 p^2 \left(m_{\eta}^2-m_{\pi}^2\right) \left(m_{\eta}^2+m_{\pi}^2\right)}{F_{\pi} f_a m_{\eta}^2}
-\frac{4 L_5 p^2 \left(m_{\eta}^2-m_{\pi}^2\right)}{\sqrt{3} F_{\pi} f_a}
\\
&-\frac{4 m_{\pi}^2 (3 L_7+L_8) \left(3 m_{\eta}^4-4 m_{\eta}^2 m_{\pi}^2+m_{\pi}^4\right)}{\sqrt{3} F_{\pi} f_a m_{\eta}^2}
-\frac{m_{\pi}^2 \left(m_{\pi}^2-3 m_{\eta}^2\right) A_0(m_{\pi})}{64 \sqrt{3} \pi ^2 F_{\pi} f_a m_{\eta}^2}
\\
&+\frac{\left[-6 p^2 \left(m_{\eta}^2-m_{\pi}^2\right)-3 m_{\eta}^2 m_{\pi}^2+m_{\pi}^4\right]A_0(m_K)}{96 \sqrt{3} \pi ^2 F_{\pi} f_a  m_{\eta}^2}
+\frac{m_{\pi}^2 \left(m_{\pi}^2-3 m_{\eta}^2\right) A_0(m_{\eta})}{192 \sqrt{3} \pi ^2 F_{\pi} f_am_{\eta}^2}
\\
&+\frac{m_a^2 \left(m_{\eta}^2-m_{\pi}^2\right)}{\sqrt{3} m_{\eta}^2 \left(m_a^2-m_{\eta}^2\right)}
\Bigg\{
\frac{6 L_4 p^2 \left(m_{\eta}^2+m_{\pi}^2\right)}{F_{\pi} f_a}
+\frac{4 L_5 m_{\eta}^2 p^2}{F_{\pi} f_a}
-\frac{12 L_6 m_{\eta}^2 \left(m_{\eta}^2+m_{\pi}^2\right)}{F_{\pi} f_a}
\\
&-\frac{12 L_7 \left(m_{\eta}^2-m_{\pi}^2\right)^2}{F_{\pi} f_a}
-\frac{4 L_8 \left(3 m_{\eta}^4-2 m_{\eta}^2 m_{\pi}^2+m_{\pi}^4\right)}{F_{\pi} f_a}
-\frac{m_{\pi}^2 A_0(m_{\pi})}{64 \pi ^2 F_{\pi} f_a }
\\
&+\frac{\left(-9 m_{\eta}^2+12 m_{K}^2+m_{\pi}^2+12 p^2\right)A_0(m_{K})}{384 \pi ^2 F_{\pi} f_a }
+\frac{\left(m_{\pi}^2-4 m_{\eta}^2\right)A_0(m_{\eta})}{192 \pi ^2 F_{\pi} f_a}
\Bigg\}\,,
\end{aligned}
\end{equation}
\begin{equation}
\begin{aligned}
\Sigma_{aa}^{(4)}(p^2)=&\frac{2 L_6 m_{\pi}^2 \left(3 m_{\eta}^2-m_{\pi}^2\right) \left(m_{\eta}^2+m_{\pi}^2\right)}{f_a^2 m_{\eta}^2}
+\frac{2  (3 L_7+L_8) m_{\pi}^4 \left(3 m_{\eta}^2-m_{\pi}^2\right)^2}{3 f_a^2 m_{\eta}^4}
\\
&+\frac{m_{\pi}^2 \left(3 m_{\eta}^2-m_{\pi}^2\right)^2 A_0(m_{\pi})}{384 \pi ^2 f_a^2 m_{\eta}^4}
+\frac{m_{\pi}^2 \left(3 m_{\eta}^2-m_{\pi}^2\right) \left(3 m_{\eta}^2+m_{\pi}^2\right) A_0(m_{K})}{576 \pi ^2 f_a^2 m_{\eta}^4}
\\
&+\frac{m_{\pi}^2 \left(3 m_{\eta}^2-m_{\pi}^2\right) \left(m_{\eta}^2+m_{\pi}^2\right) A_0(m_{\eta})}{1152 \pi ^2 f_a^2 m_{\eta}^4}
\\
&+\frac{m_a^2 \left(m_{\eta}^2-m_{\pi}^2\right)}{2 \sqrt{3} m_{\eta}^2 \left(m_a^2-m_{\eta}^2\right)}
\Bigg\{
\frac{4 \sqrt{3} L_4 p^2 \left(m_{\eta}^4-m_{\pi}^4\right)}{f_a^2 m_{\eta}^2}
+\frac{8 L_5 p^2 \left(m_{\eta}^2-m_{\pi}^2\right)}{\sqrt{3} f_a^2}
\\
&+\frac{8(3 L_7+L_8)m_{\pi}^2\left(3 m_{\eta}^4-4 m_{\eta}^2 m_{\pi}^2+m_{\pi}^4\right)}{\sqrt{3} f_a^2 m_{\eta}^2}
+\frac{\left(m_{\pi}^4-3 m_{\eta}^2 m_{\pi}^2\right) A_0(m_{\pi})}{32 \sqrt{3} \pi ^2 f_a^2 m_{\eta}^2}
\\
&+\frac{\left[3 m_{\eta}^2 \left(m_{\pi}^2+2 p^2\right)-m_{\pi}^2 \left(m_{\pi}^2+6 p^2\right)\right]A_0(m_{K})}{48 \sqrt{3} \pi ^2 f_a^2 m_{\eta}^2}
+\frac{m_{\pi}^2 \left(3 m_{\eta}^2-m_{\pi}^2\right) A_0(m_{\eta})}{96 \sqrt{3} \pi ^2 f_a^2 m_{\eta}^2}
\Bigg\}
\\
&+\frac{m_a^4 \left(m_{\eta}^2-m_{\pi}^2\right)^2}{12 m_{\eta}^4 \left(m_a^2-m_{\eta}^2\right)^2}
\Bigg\{
\frac{24 L_6 m_{\eta}^2 \left(m_{\eta}^2+m_{\pi}^2\right)}{f_a^2}
-\frac{12 L_4 p^2 \left(m_{\eta}^2+m_{\pi}^2\right)}{f_a^2}
-\frac{8 L_5 m_{\eta}^2 p^2}{f_a^2}
\\
&+\frac{24 L_7 \left(m_{\eta}^2-m_{\pi}^2\right)^2}{f_a^2}
+\frac{8 L_8 \left(3 m_{\eta}^4-2 m_{\eta}^2 m_{\pi}^2+m_{\pi}^4\right)}{f_a^2}
+\frac{m_{\pi}^2 A_0(m_{\pi})}{32 \pi ^2 f_a^2}
\\
&-\frac{\left(m_{\pi}^2+3 p^2\right) A_0(m_{K})}{48 \pi ^2 f_a^2}
-\frac{\left(m_{\pi}^2-4 m_{\eta}^2\right) A_0(m_{\eta})}{96 \pi ^2 f_a^2}
\Bigg\}\,,
\end{aligned}
\end{equation}
where 
\begin{equation}
A_0(m)=-(\mu^2)^{\frac{d-4}{2}} m^2 \left(R+\log\frac{m^2}{\mu ^2}\right)\,,
\quad
R=\frac{2}{d-4}-\Gamma'(1)-\log(4\pi)-1\,,
\end{equation}
with $d\to4$. The UV divergence in the loop corrections is renormalized in the physical quantities with the bare LECs~\cite{Gasser:1984gg}
\begin{equation}
L_i=L_i^r(\mu)+\Gamma_i\frac{(\mu^2)^{\frac{d-4}{2}}}{2(4\pi)^2}R\,,
\end{equation}
with $\Gamma_1=3/32$, $\Gamma_2=3/16$, $\Gamma_3=0$, $\Gamma_4=1/8$, $\Gamma_5=3/8$, $\Gamma_6=11/144$, $\Gamma_7=0$, $\Gamma_8=5/48$.

The full perturbative $\chi$PT amplitude of the $\eta\to\pi\pi a$ process is given by
\begin{equation}\label{eq.decayampfull}
\mathcal{M}_{\eta;\pi\pi a}^{(2+4)}(s,t,u)= \mathcal{M}_{\eta;\pi\pi a}^{(2)}(s,t,u)+\mathcal{M}_{\eta;\pi\pi a}^{(4)}(s,t,u),
\end{equation}
where the LO part $\mathcal{M}_{\eta;\pi\pi a}^{(2)}(s,t,u)$ is given in Eq.~\eqref{eq.MLO} and the NLO part will be decomposed into four pieces: the $s$-channel loop diagram (Fig.~\ref{fig.1d}) contribution $\mathcal{M}_{\eta;\pi\pi a}^{s}$, the $t$- and $u$-channel loop diagrams (Figs.~\ref{fig.1e} and \ref{fig.1f}) contributions $\mathcal{M}_{\eta;\pi\pi a}^{t,u}$, and the $\mathcal{M}_{\eta;\pi\pi a}^{\mathrm{c+tad}}$ term that includes the contributions from the $\mathcal{O}(p^4)$ contact diagram (Fig.~\ref{fig.1b}), tadpole diagram (Fig.~\ref{fig.1c}), the wave function renormalizations of $\pi$ and $\eta$, and the NLO ALP-$\eta$ mixing, as well as the correction of Eq.~(\ref{eq.CorrectionFromLOPara}). The explicit expression for the NLO part is 
\begin{equation}
\mathcal{M}_{\eta;\pi\pi a}^{(4)}= \mathcal{M}_{\eta;\pi\pi a}^{s}+\mathcal{M}_{\eta;\pi\pi a}^{t}+\mathcal{M}_{\eta;\pi\pi a}^{u}+\mathcal{M}_{\eta;\pi\pi a}^{\mathrm{c}+\mathrm{tad}}\,,
\label{eq.decayampp4}
\end{equation}
where  
\begin{align}
&\mathcal{M}_{\eta;\pi\pi a}^{s}=\frac{m_{\eta}^2\left[9 s \left(3 s-m_a^2\right)-48 m_{\pi}^4+81 m_{\pi}^2 s\right]}{2304 \sqrt{3} \pi ^2 f_a F_{\pi}^3 m_{\eta}^2}
-\frac{27 m_{\eta}^4 s-16 m_{\pi}^6+18 m_{\pi}^4 s}{2304 \sqrt{3} \pi ^2 f_a F_{\pi}^3 m_{\eta}^2}
\notag\\
&+\frac{9 m_{\pi}^2 s \left(m_a^2-3 s\right)}{2304 \sqrt{3} \pi ^2 f_a F_{\pi}^3 m_{\eta}^2}
+\frac{\left(3 m_{\eta}^2-m_{\pi}^2\right) \left(7 m_{\pi}^2-6 s\right) \mu_{\pi}}{18 \sqrt{3} f_a F_{\pi} m_{\eta}^2}
-\frac{m_{\pi}^4 \left(m_{\pi}^2-3 m_{\eta}^2\right) \mu_{\eta}}{18 \sqrt{3} f_a F_{\pi} m_{\eta}^4}
\notag\\
&+\frac{\left(2 m_{K}^2-3 s\right) \left[3 m_{\eta}^2 \left(m_{\pi}^2+3 s\right)-3 m_a^2 \left(m_{\eta}^2-m_{\pi}^2\right)-9 m_{\eta}^4+2 m_{\pi}^4-9 m_{\pi}^2 s\right]\mu_{K}}{72 \sqrt{3} f_a F_{\pi} m_{\eta}^2 m_{K}^2}
\notag\\
&+\frac{m_{\pi}^2 \left(m_{\pi}^2-3 m_{\eta}^2\right) \left(m_{\pi}^2-2 s\right) \mathcal{F}_{B_0}(s,m_{\pi},m_{\pi})}{192 \sqrt{3} \pi ^2 f_a F_{\pi}^3 m_{\eta}^2}
+\frac{m_{\pi}^4 \left(m_{\pi}^2-3 m_{\eta}^2\right) \mathcal{F}_{B_0}(s,m_{\eta},m_{\eta})}{576 \sqrt{3} \pi ^2 f_a F_{\pi}^3 m_{\eta}^2}
\notag\\
&-\frac{s\left[3 m_a^2 \left(m_{\eta}^2-m_{\pi}^2\right)+9 m_{\eta}^4-3 m_{\eta}^2 \left(m_{\pi}^2+3 s\right)-2 m_{\pi}^4+9 m_{\pi}^2 s\right]\mathcal{F}_{B_0}(s,m_{K},m_{K})}{768 \sqrt{3} \pi ^2 f_a F_{\pi}^3 m_{\eta}^2}
\notag\\
&+\frac{m_a^2}{m_a^2-m_{\eta}^2}\Bigg\{\frac{\left(m_{\eta}^2-m_{\pi}^2\right)}{4608 \sqrt{3} \pi ^2 f_a F_{\pi}^3 m_{\eta}^2}
\Big[18 s\left(m_a^2+2 m_{K}^2-3 s\right)-m_{\eta}^2\left(32 m_{\pi}^2+9 s\right)+32 m_{\pi}^4
\notag\\
&-45 m_{\pi}^2 s\Big]
-\frac{\left(m_{\eta}^2-m_{\pi}^2\right) \left(7 m_{\pi}^2-6 s\right) \mu_{\pi}}{18 \sqrt{3} f_a F_{\pi} m_{\eta}^2}
+\frac{\left(4 m_{\eta}^4 m_{\pi}^2-5 m_{\eta}^2 m_{\pi}^4+m_{\pi}^6\right) \mu_{\eta}}{18 \sqrt{3} f_a F_{\pi} m_{\eta}^4}
\notag\\
&+\frac{\left(m_{\eta}^2-m_{\pi}^2\right) \mu_{K}}{144 \sqrt{3} f_a F_{\pi} m_{\eta}^2 m_{K}^2}
\Big[3 s \left(3 m_{\eta}^2-m_{\pi}^2+18 s\right)+4 m_{K}^2 \left(3 m_{\eta}^2+2 m_{\pi}^2-24 s\right)
\notag\\
&+6 m_a^2 \left(2 m_{K}^2-3 s\right)\Big]
+\frac{m_{\pi}^2 \left(m_{\eta}^2-m_{\pi}^2\right) \left(m_{\pi}^2-2 s\right) \mathcal{F}_{B_0}(s,m_{\pi},m_{\pi})}{192 \sqrt{3} \pi ^2 f_a F_{\pi}^3 m_{\eta}^2}
\notag\\
&-\frac{m_{\pi}^2\left(4 m_{\eta}^4-5 m_{\eta}^2 m_{\pi}^2+m_{\pi}^4\right)\mathcal{F}_{B_0}(s,m_{\eta},m_{\eta})}{576 \sqrt{3} \pi ^2 f_a F_{\pi}^3 m_{\eta}^2}
-\frac{s \left(m_{\eta}^2-m_{\pi}^2\right) \mathcal{F}_{B_0}(s,m_{K},m_{K})}{1536 \sqrt{3} \pi ^2 f_a F_{\pi}^3 m_{\eta}^2}
\notag\\
&\times\big(18 s-6 m_a^2+3 m_{\eta}^2-12 m_{K}^2-m_{\pi}^2\big)\Bigg\}\,,
\end{align}

\begin{align}
&\mathcal{M}_{\eta;\pi\pi a}^{t}=\frac{3 m_a^2 \left(m_{\eta}^2-m_{\pi}^2\right) \left(3 m_{\eta}^2+m_{\pi}^2-3 t\right)+3 m_{\eta}^2 \left(2 m_{\pi}^4+15 m_{\pi}^2 t+9 t^2\right)}{2304 \sqrt{3} \pi ^2 f_a F_{\pi}^3 m_{\eta}^2}
\notag\\
&-\frac{27 m_{\eta}^4 \left(m_{\pi}^2+t\right)}{2304 \sqrt{3} \pi ^2 f_a F_{\pi}^3 m_{\eta}^2}
-\frac{11 m_{\pi}^6-18 m_{\pi}^4 t+27 m_{\pi}^2 t^2}{2304 \sqrt{3} \pi ^2 f_a F_{\pi}^3 m_{\eta}^2}
+\frac{m_{\pi}^2 \left(m_{\pi}^2-3 m_{\eta}^2\right) \mu_{\pi}}{18 \sqrt{3} f_a F_{\pi} m_{\eta}^2}
\notag\\
&+\frac{m_{\pi}^4 \left(m_{\pi}^2-3 m_{\eta}^2\right) \mu_{\eta}}{18 \sqrt{3} f_a F_{\pi} m_{\eta}^4}
+\frac{\left(3 m_{\eta}^2+2 m_{K}^2+m_{\pi}^2-3 t\right)\mu_{K}}{24 \sqrt{3} f_a F_{\pi} m_{\eta}^2 m_{K}^2}
\Big[m_a^2\left(m_{\pi}^2-m_{\eta}^2\right)+m_{\pi}^4
\notag\\
&+3 m_{\eta}^2 \left(m_{\pi}^2+t\right)-3 m_{\pi}^2 t\Big]
-\frac{m_{\pi}^4 \left(3 m_{\eta}^4-4 m_{\eta}^2 m_{\pi}^2+m_{\pi}^4\right)}{576 \sqrt{3} \pi ^2 f_a F_{\pi}^3 m_{\eta}^2 t} \log\frac{m_{\eta}^2}{m_{\pi}^2}
-\frac{m_{\pi}^4 \left(m_{\pi}^2-3 m_{\eta}^2\right)}{288 \sqrt{3} \pi ^2 f_a F_{\pi}^3 m_{\eta}^2}
\notag\\
&\times\mathcal{F}_{B_0}(t,m_{\eta},m_{\pi})
-\frac{\left(3 m_{\eta}^2+m_{\pi}^2-3 t\right)\mathcal{F}_{B_0}(t,m_{K},m_{K})}{768 \sqrt{3} \pi ^2 f_a F_{\pi}^3 m_{\eta}^2}
\Big[m_a^2 \left(m_{\pi}^2-m_{\eta}^2\right)+m_{\pi}^4
\notag\\
&+3 m_{\eta}^2 \left(m_{\pi}^2+t\right)-3 m_{\pi}^2 t\Big]
+\frac{m_a^2}{m_a^2-m_{\eta}^2}\Bigg\{
\frac{\left(m_{\pi}^2-m_{\eta}^2\right)}{2304 \sqrt{3} \pi ^2 f_a F_{\pi}^3 m_{\eta}^2}
\Big[3 m_a^2 \left(3 m_{\eta}^2+m_{\pi}^2-3 t\right)
\notag\\
&+18 m_{\eta}^4+15 m_{\eta}^2 \left(m_{\pi}^2-3 t\right)+11 m_{\pi}^4-18 m_{\pi}^2 t+27 t^2\Big]
+\frac{m_{\pi}^2 \left(m_{\eta}^2-m_{\pi}^2\right) \mu_{\pi}}{18 \sqrt{3} f_a F_{\pi} m_{\eta}^2}
\notag\\
&+\frac{m_{\pi}^4 \left(m_{\eta}^2-m_{\pi}^2\right) \mu_{\eta}}{18 \sqrt{3} f_a F_{\pi} m_{\eta}^4}
+\frac{\left(m_{\eta}^2-m_{\pi}^2\right) \mu_{K}}{24 \sqrt{3} f_a F_{\pi} m_{\eta}^2 m_{K}^2}
\Big[m_a^2 \left(3 m_{\eta}^2+2 m_{K}^2+m_{\pi}^2-3 t\right)+6 m_{\eta}^4+5 m_{\eta}^2 m_{\pi}^2
\notag\\
&-15 m_{\eta}^2 t+2 m_{K}^2 \left(5 m_{\eta}^2+2 m_{\pi}^2-5 t\right)+m_{\pi}^4-6 m_{\pi}^2 t+9 t^2\Big]
+\frac{m_{\pi}^4 \left(m_{\eta}^2-m_{\pi}^2\right)^2}{576 \sqrt{3} \pi ^2 f_a F_{\pi}^3 m_{\eta}^2 t} \log\frac{m_{\eta}^2}{m_{\pi}^2}
\notag\\
&+\frac{m_{\pi}^4 \left(m_{\pi}^2-m_{\eta}^2\right) \mathcal{F}_{B_0}(t,m_{\eta},m_{\pi})}{288 \sqrt{3} \pi ^2 f_a F_{\pi}^3 m_{\eta}^2}
-\frac{\left(m_{\eta}^2-m_{\pi}^2\right) \left(3 m_{\eta}^2+m_{\pi}^2-3 t\right) \mathcal{F}_{B_0}(t,m_{K},m_{K})}{768 \sqrt{3} \pi ^2 f_a F_{\pi}^3 m_{\eta}^2}
\notag\\
&\times\left(m_a^2+2 m_{\eta}^2+m_{\pi}^2-3 t\right)\Bigg\}\,,
\end{align}

\begin{equation}
\mathcal{M}_{\eta;\pi\pi a}^{u}=\mathcal{M}_{\eta;\pi\pi a}^{t}\big|_{t\to u}\,,
\end{equation}

\begin{align}
&\mathcal{M}_{\eta;\pi\pi a}^{\mathrm{c}+\mathrm{tad}}=-\frac{4 L_1^r \left(m_{\eta}^2-m_{\pi}^2\right) \left(s-2 m_{\pi}^2\right) \left(m_a^2+m_{\eta}^2-s\right)}{\sqrt{3} f_a F_{\pi}^3 m_{\eta}^2}
-\frac{2 L_2^r \left(m_{\eta}^2-m_{\pi}^2\right) }{\sqrt{3} f_a F_{\pi}^3 m_{\eta}^2}
\Big[m_a^4+m_{\eta}^4
\notag\\
&-s\left(m_a^2+m_{\eta}^2\right)-2 m_{\pi}^4+2 m_{\pi}^2 (t+u)-t^2-u^2\Big]
+\frac{2 L_3^r \left(m_{\eta}^2-m_{\pi}^2\right) }{3 \sqrt{3} f_a F_{\pi}^3 m_{\eta}^2}
\Big(s^2+t^2+u^2-m_a^4
\notag\\
&-m_{\eta}^4-2 m_{\pi}^4\Big)
+\frac{4 L_4^r m_{\pi}^2 \left(m_{\pi}^2-m_{\eta}^2\right) \left(m_a^2+m_{\eta}^2-s\right)}{\sqrt{3} f_a F_{\pi}^3 m_{\eta}^2}
+\frac{2 L_5^r m_{\pi}^2 }{3 \sqrt{3} f_a F_{\pi}^3 m_{\eta}^2}
\Big[m_a^2 \left(m_{\eta}^2+m_{\pi}^2\right)
\notag\\
&-6 m_{\eta}^4+m_{\eta}^2 m_{\pi}^2-m_{\pi}^4\Big]
-\frac{4 L_7^r \left(6 m_{\eta}^6 m_{\pi}^2-13 m_{\eta}^4 m_{\pi}^4+2 m_{\eta}^2 m_{\pi}^6+m_{\pi}^8\right)}{\sqrt{3} f_a F_{\pi}^3 m_{\eta}^4}
-\frac{4 L_8^r m_{\pi}^4 }{3 \sqrt{3} f_a F_{\pi}^3 m_{\eta}^4}
\notag\\
&\times\big(2 m_{\eta}^2 m_{\pi}^2+m_{\pi}^4-15 m_{\eta}^4\big)
+\frac{\left(13 m_{\eta}^2 m_{\pi}^4+3 m_{\pi}^6-30 m_{\eta}^4 m_{\pi}^2\right) \mu_{\pi}}{18 \sqrt{3} f_a F_{\pi} m_{\eta}^4}
-\frac{\left(m_{\pi}^6-3 m_{\eta}^2 m_{\pi}^4\right) \mu_{\eta}}{18 \sqrt{3} f_a F_{\pi} m_{\eta}^4}
\notag\\
&+\frac{\left[18 m_a^2 \left(m_{\eta}^4-m_{\eta}^2 m_{\pi}^2\right)+5 \left(3 m_{\eta}^4 m_{\pi}^2-7 m_{\eta}^2 m_{\pi}^4-2 m_{\pi}^6\right)\right]\mu_{K}}{90 \sqrt{3} f_a F_{\pi} m_{\eta}^4}
\notag\\
&+\frac{m_a^2}{m_a^2-m_{\eta}^2}\Bigg\{
\frac{4 L_1^r \left(m_{\eta}^2-m_{\pi}^2\right) \left(s-2 m_{\pi}^2\right) \left(m_a^2+m_{\eta}^2-s\right)}{\sqrt{3} f_a F_{\pi}^3 m_{\eta}^2}
+\frac{2 L_2^r \left(m_{\eta}^2-m_{\pi}^2\right)}{\sqrt{3} f_a F_{\pi}^3 m_{\eta}^2}
\Big[m_a^4+m_{\eta}^4
\notag\\
&-s \left(m_a^2+m_{\eta}^2\right)-2 m_{\pi}^4+2 m_{\pi}^2 (t+u)-t^2-u^2\Big]
-\frac{2 L_3^r \left(m_{\eta}^2-m_{\pi}^2\right)}{3 \sqrt{3} f_a F_{\pi}^3 m_{\eta}^2}
\Big(s^2+t^2+u^2-m_a^4
\notag\\
&-m_{\eta}^4-2 m_{\pi}^4\Big)
+\frac{4 L_4^r \left(m_{\eta}^2-m_{\pi}^2\right) \left[m_a^2 m_{\pi}^2+m_{\eta}^2 \left(3 m_{\pi}^2-s\right)-m_{\pi}^2 s\right]}{\sqrt{3} f_a F_{\pi}^3 m_{\eta}^2}
+\frac{2 L_5^r m_{\pi}^2 \left(m_{\eta}^2-m_{\pi}^2\right)}{3 \sqrt{3} f_a F_{\pi}^3 m_{\eta}^2}
\notag\\
&\times\left(m_a^2+2 m_{\eta}^2-m_{\pi}^2\right)
+\frac{16 L_6^r \left(m_{\pi}^4-m_{\eta}^2 m_{\pi}^2\right)}{\sqrt{3} f_a F_{\pi}^3}
+\frac{4 L_7^r \left(4 m_{\eta}^2+m_{\pi}^2\right) \left(m_{\pi}^3-m_{\eta}^2 m_{\pi}\right)^2}{\sqrt{3} f_a F_{\pi}^3 m_{\eta}^4}
\notag\\
&+\frac{4 L_8^r m_{\pi}^4 \left(2 m_{\eta}^2 m_{\pi}^2+m_{\pi}^4-3 m_{\eta}^4\right)}{3 \sqrt{3} f_a F_{\pi}^3 m_{\eta}^4}
+\frac{\left(10 m_{\eta}^4 m_{\pi}^2-13 m_{\eta}^2 m_{\pi}^4-3 m_{\pi}^6\right) \mu_{\pi}}{18 \sqrt{3} f_a F_{\pi} m_{\eta}^4}
+\frac{\left(m_{\pi}^6-3 m_{\eta}^2 m_{\pi}^4\right)}{18 \sqrt{3} f_a F_{\pi} m_{\eta}^4}
\notag\\
&\times\mu_{\eta}
+\frac{\left[7 m_{\eta}^4 m_{\pi}^2+50 m_{\eta}^2 m_{\pi}^4+10 m_{\pi}^6-3 m_a^2 \left(m_{\eta}^4-m_{\eta}^2 m_{\pi}^2\right)-27 m_{\eta}^6\right]\mu_{K}}{90 \sqrt{3} f_a F_{\pi} m_{\eta}^4}
\Bigg\}
\notag\\
&+\frac{2 L_4^r m_a^2 m_{\pi}^2 \left(m_{\pi}^4-m_{\eta}^4\right)}{\sqrt{3} f_a F_{\pi}^3 m_{\eta}^2 \left(m_a^2-m_{\eta}^2\right)}
+\frac{4 L_5^r m_a^2 m_{\pi}^2 \left(m_{\pi}^2-m_{\eta}^2\right)}{3 \sqrt{3} f_a F_{\pi}^3 \left(m_a^2-m_{\eta}^2\right)}
+\frac{4 m_{\pi}^4 (3 L_7^r+L_8^r)}{3 \sqrt{3} f_a F_{\pi}^3 m_{\eta}^2 \left(m_{\eta}^2-m_a^2\right)}
\Big(3 m_{\eta}^4+m_{\pi}^4
\notag\\
&-4 m_{\eta}^2 m_{\pi}^2\Big)
-\frac{m_{\pi}^4 \left(m_{\pi}^2-3 m_{\eta}^2\right) \mu_{\pi}}{6 \sqrt{3} f_a F_{\pi} m_{\eta}^2 \left(m_{\eta}^2-m_a^2\right)}
+\frac{m_{\pi}^4 \left(m_{\pi}^2-3 m_{\eta}^2\right) \mu_{\eta}}{18 \sqrt{3} f_a F_{\pi} m_{\eta}^2 \left(m_{\eta}^2-m_a^2\right)}
+\Big[m_{\pi}^4-3 m_{\eta}^2 m_{\pi}^2
\notag\\
&-6 m_a^2\left(m_{\eta}^2-m_{\pi}^2\right)\Big]
\frac{m_{\pi}^2 \mu_{K} }{9 \sqrt{3} f_a F_{\pi} m_{\eta}^2 \left(m_{\eta}^2-m_a^2\right)}
+\frac{m_a^2}{m_a^2-m_{\eta}^2}\Bigg\{
\frac{2 L_4^r m_{\pi}^2 \left(m_{\eta}^4-m_{\pi}^4\right)}{\sqrt{3} f_a F_{\pi}^3 m_{\eta}^2}
\notag\\
&+\frac{4 L_5^r m_{\pi}^2 \left(m_{\eta}^2-m_{\pi}^2\right)}{3 \sqrt{3} f_a F_{\pi}^3}
+\frac{\left(m_{\pi}^4-m_{\eta}^2 m_{\pi}^2\right) \mu_{K}}{3 \sqrt{3} f_a F_{\pi} m_{\eta}^2}
\Bigg\}\,,
\end{align}

with 
\begin{align}
\mu_p=\frac{m_p^2}{2 (4 \pi F_{\pi})^2}\log\frac{m_p^2}{\mu^2}\,,
\end{align}
\begin{equation}
\mathcal{F}_{B_0}(s,m_1,m_2)=\frac{\sqrt{\lambda(s,m_1^2,m_2^2)}}{s}
\log\left(\frac{m_1^2+m_2^2-s+\sqrt{\lambda(s,m_1^2,m_2^2)}}{2m_1m_2}\right)\,.
\end{equation}
In the region of $s>(m_1+m_2)^2$, the function $\mathcal{F}_{B_0}(s,m_1,m_2)$ develops the unitary cut, which is given by 
\begin{equation}
\mathrm{Im}\mathcal{F}_{B_0}(s+i0^+,m_1,m_2)=\theta\left[s-(m_1+m_2)^2\right]\pi\frac{\sqrt{\lambda(s,m_1^2,m_2^2)}}{s}\,.
\end{equation}

By assuming the lowest resonance saturation to the $\mathcal{O}(p^4)$ LECs and the large $N_C$ relations of the resonance couplings in Ref.~\cite{Ecker:1988te}
\begin{equation}
\begin{aligned}
&L_1^r = \frac{G_V^2}{8 M_V^2}\,,
\quad
L_2^r = \frac{G_V^2}{4 M_V^2}\,,
\quad
L_3^r = \frac{c_d^2}{2 M_S^2}-\frac{3 G_V^2}{4 M_V^2}\,,
\quad
L_4^r = 0\,,
\\
&L_5^r = \frac{c_d c_m}{M_S^2}\,,
\quad
L_6^r = 0\,,
\quad
L_7^r = 0\,,
\quad
L_8^r = \frac{c_m^2}{2 M_S^2}-\frac{d_m^2}{2 M_P^2}\,,
\end{aligned}
\end{equation}
the $L_{i=1,\cdots,8}^r$ terms in Eq.~\eqref{eq.decayampp4} become 
\begin{equation}
\begin{aligned}
\mathcal{M}_{\eta;\pi\pi a}^{\mathrm{LECs}}=
&\frac{c_d^2 \left(m_{\eta}^2-m_{\pi}^2\right) \left(m_a^4+m_{\eta}^4+2 m_{\pi}^4-s^2-t^2-u^2\right)}{3 \sqrt{3} f_a F_{\pi}^3 M_S^2 \left(m_a^2-m_{\eta}^2\right)}
+\frac{4 c_m^2 m_{\pi}^4 \left(2 m_a^2-3 m_{\eta}^2+m_{\pi}^2\right)}{\sqrt{3} f_a F_{\pi}^3 M_S^2 \left(m_a^2-m_{\eta}^2\right)}
\\
&+\frac{2 c_d c_m m_{\pi}^2 \left[2 m_a^4-m_a^2 \left(5 m_{\eta}^2+3 m_{\pi}^2\right)+6 m_{\eta}^4-m_{\eta}^2 m_{\pi}^2+m_{\pi}^4\right]}{3 \sqrt{3} f_a F_{\pi}^3 M_S^2 \left(m_a^2-m_{\eta}^2\right)}
\\
&-\frac{4 d_m^2 m_{\pi}^4 \left(2 m_a^2-3 m_{\eta}^2+m_{\pi}^2\right)}{\sqrt{3} f_a F_{\pi}^3 M_P^2 \left(m_a^2-m_{\eta}^2\right)}\,,
\end{aligned}
\end{equation}
which shows that the vector resonances do not contribute to the $\eta\to\pi\pi a$ process.

\bibliography{ALP_in_eta_decay}
\bibliographystyle{apsrev4-2}

\end{document}